\documentclass[12pt,preprint]{aastex}

\newcommand{\ace}{\alpha_{\rm CE}}
\newcommand{\Hubble}{{\it Hubble Space Telescope}}
\newcommand{\HST}{{\it HST}}
\newcommand{\Ibanoglu}{\.Ibano\u{g}lu}
\newcommand{\IUE}{{\it IUE}}
\newcommand{\Lya}{Ly$\alpha$}
\newcommand{\Teff}{T_{\rm eff}}
\newcommand{\VTau}{V471~Tau}
\newcommand{\logg}{\log g}
\newcommand{\kms}{{\>\rm km\>s^{-1}}}

\newcommand{\kd}{K~dwarf}

\shorttitle{{\it HST\/} Spectroscopy of V471 Tauri}
\shortauthors{O'Brien et al.}

\begin{document}

\title{\Hubble\/ Spectroscopy of V471 Tauri: \\
Oversized K Star, Paradoxical White Dwarf\altaffilmark{1} }
\author{M. Sean O'Brien and Howard E. Bond}
\affil{Space Telescope Science Institute, 
		 3700 San Martin Dr.,
		 Baltimore, MD 21218; 
		 obrien@stsci.edu, 
		 bond@stsci.edu}
\and
\author{Edward M. Sion}
\affil{Department of Astronomy \& Astrophysics, 
		 Villanova University,
		 Villanova, PA 19085;
		 edward.sion@villanova.edu}

\altaffiltext{1}{Based on observations with  the NASA/ESA {\it Hubble Space
Telescope}, obtained at the Space Telescope Science Institute, which is
operated by AURA, Inc., under NASA contract NAS5-26555.}

\begin{abstract}

We have used the Goddard High Resolution Spectrograph onboard the {\it Hubble
Space Telescope\/} to obtain Lyman-$\alpha$ spectra of the hot white-dwarf (WD)
component of the short-period eclipsing DA+dK2 pre-cataclysmic binary
V471~Tauri, a member of the Hyades star cluster. 

Radial velocities of the WD were determined from eight post-COSTAR spectra,
obtained near the two quadratures of the orbit. When combined with ground-based
measurements of the dK velocities, eclipse timings, and a determination of the
dK star's rotational velocity, the data constrain the orbital inclination to be
$i=77^\circ$, and yield dynamical masses for the components of $M_{\rm
WD}=0.84$ and $M_{\rm dK}=0.93\,M_\sun$. Model-atmosphere fitting of the
Ly$\alpha$ profile provides the effective temperature (34,500~K) and surface
gravity ($\log g=8.3$) of the WD\null. 

The radius of the dK component is about 18\% larger than that of a normal
Hyades dwarf of the same mass. This expansion is attributed to the large degree
of coverage of the stellar surface by starspots, which is indicated both by
radiometric measurements and ground-based Doppler imaging; in response, the
star has expanded in order to maintain the luminosity of a $0.93\,M_\sun$
dwarf.  

The radius of the WD, determined from a radiometric analysis and from eclipse
ingress timings, is $0.0107\,R_\sun$. The position of the star in the
mass-radius plane is in full accord with theoretical predictions for a
degenerate carbon-oxygen WD with a surface temperature equal to that observed.
The position of the WD in the H-R diagram is also fully consistent with that
expected for a WD with our dynamically measured mass.  Both comparisons with
theory are probably the most stringent yet made for any WD\null. The theoretical
cooling age of the WD is $10^7$~yr.

The high effective temperature and high mass of the WD present an evolutionary
paradox. The WD is the most massive one known in the Hyades, but also the
hottest and youngest, in direct conflict with expectation. We examine possible
resolutions of the paradox, including the possibility of a nova outburst in the
recent past, but conclude that the most likely explanation is that the WD is
indeed very young, and is descended from a blue straggler. A plausible scenario
is that the progenitor system was a triple, with a close inner pair of
main-sequence stars whose masses were both similar to that of the present
cluster turnoff. These stars became an Algol-type binary, which merged after
several hundred million years to produce a single blue straggler of about twice
the turnoff mass. When this star evolved to the AGB phase, it underwent a
common-envelope interaction with a distant dK companion, which spiralled down
to its present separation, and ejected the envelope. We estimate that the
common-envelope efficiency parameter, $\ace$, was of order 0.3--1.0, in good
agreement with recent hydrodynamical simulations.

\end{abstract}


\section{Introduction}

V471 Tauri (BD~+16$^\circ$516) is a remarkable binary system belonging to the
Hyades star cluster.  It is an eclipsing binary with an orbital period of only
12.5~hr, whose components are a hot DA white dwarf (WD) and a dK2 main-sequence
(MS) star.  Following discovery of the eclipses by \citet{nel70}, 
\citet{vau72} pointed out that \VTau's current system mass is much lower than
the mass that the progenitor system must have had. It was also realized during
the early 1970's that the system's current orbital separation is much smaller
than the radius of the red giant that had to have existed in the binary in
order to form the WD.

These puzzles would be resolved if the progenitor binary had an initial orbital
separation large enough for the original primary star to attain red-giant
dimensions before encountering its MS companion. Only then did
extensive loss of mass and orbital angular momentum lead to extreme contraction
of the orbit.  In a classic paper, \citet{pac76} proposed a scenario in which
the expanding giant primary engulfs its companion in a ``common envelope''
(CE)\null.  Friction then leads to a rapid  spiraling-down of the orbit, during
which coalescence of the red giant's core and the MS companion may
be avoided if the CE  can be spun up to the critical rotation velocity and
ejected from the system.  Planetary nebulae  with extremely close binary
nuclei \citep[and references therein]{bon90,bon00} provide the most
direct evidence that CE interactions and envelope ejections do occur in
nature.  As discussed by Paczy\'nski (1976), V471~Tau is most probably directly
descended from a system of this type, and will evolve eventually into a
cataclysmic variable (CV); thus V471~Tau is often considered to be the
prototypical ``pre-cataclysmic'' binary (e.g., Bond 1985).

A critical parameter that determines the outcome of a CE interaction is the
efficiency with which orbital energy is converted into ejection of
matter  from the system. This efficiency, denoted $\ace$, is defined (see
Iben \& Livio 1993) by
\begin{equation}
\alpha_{\rm CE} = \frac{\Delta E_{\rm{bind}}}{\Delta E_{\rm{orb}}} \, ,
\end{equation}
where $\Delta E_{\rm{bind}}$ is the binding energy of the ejected material and
$\Delta E_{\rm{orb}}$ is the change in the orbital energy of the binary between
the beginning and end of the spiraling-in process.  


If $\alpha_{\rm CE}$ is high, $\sim$1, then envelope ejection occurs early in
the interaction, and most of the binaries emerging from CEs will still have
relatively long orbital periods (typically 10--100 days; cf.\ Yungelson,
Tutukov, \& Livio 1993). If, however, $\ace$ is much lower than unity, then
short-period CE descendants along with mergers will be much more common, making
it easier to produce close-binary planetary nuclei, \VTau-like detached
WD/red-dwarf binaries, CVs, and eventually Type~Ia supernovae. Knowledge of the
typical values of $\ace$ is therefore crucial in understanding the evolution of
populations of binary systems.

\VTau\ is one of two known post-CE, pre-CV systems in the Hyades cluster (the 
other being HZ~9).  \VTau's cluster membership is confirmed by {\it
Hipparcos\/} proper-motion and parallax measurements 
\citep[][]{pro96,per98,deBruijne01}.  The high surface temperature of the WD,
$\Teff\simeq35,000$~K, implies a very short cooling age of  $\sim$$10^7$~yr
(see \S7.3). Thus, it ought to be possible to assume the original mass of the
WD progenitor to have been equal to the cluster's current MS turnoff mass; by
measuring the masses of both components of the current system, we would then
have all of the information needed to determine $\ace$ entirely empirically.


Since the V471~Tau system eclipses, we simply require the radial-velocity
curves (along with, for highest accuracy,  the inclination of the orbit with
respect to the line of sight) in order to measure purely dynamical masses of
both components.  However, it has not proven possible to measure reliable radial
velocities of the WD component from the ground \citep[e.g.,][]{gil83}, since
its optical light is badly swamped by that of the K2 component. Thus, the 
primary goal of this project was to measure radial velocities of the  WD in the
ultraviolet, where the K2 star contributes negligibly  to the flux.

Additionally, UV spectroscopic observations of \VTau\ would allow investigation
of other astrophysical questions. The WD mass-radius relation is  still
relatively poorly constrained by  observation, with Sirius~B and Procyon~B
being virtually the only WDs with very accurately measured dynamical masses.
V471~Tau, a bright, detached eclipsing binary, offers an excellent opportunity
to add to this short list. 

Also of interest is the finding by \citet{jen86} of a periodic 9.25-min 
modulation in the soft X-ray flux of V471~Tau, subsequently found also in the
optical band by Robinson, Clemens, \& Hine (1988).   \citet{cle92} showed that
the X-ray and optical modulations are  $180^\circ$ out of phase, and argued
that they arise because of magnetic accretion from the K~dwarf's wind onto the
pole(s) of the WD\null. In this picture, the WD's rotation period is 9.25~min,
and metallic  opacity darkens the pole(s) of the WD in soft X-rays while
brightening them through flux redistribution in the optical (see Dupuis et al.\
1997). Time-resolved UV spectroscopy offers the possibility of testing this
model by detecting absorption lines of accreted metals.

With the above motivations, we obtained observations of \VTau\ with the Goddard
High-Resolution Spectrograph (GHRS) onboard the \Hubble\ (\HST\/)\null. This
paper deals with the measurement and interpretation of the radial-velocity
variations of the WD, and use of the GHRS spectra to determine the temperature
and gravity of the WD\null. In addition, we have used the same set of
observations to detect a Zeeman-split photospheric \ion{Si}{3} absorption line
whose strength is modulated on the rotation period of the WD, strongly
confirming the magnetic accretion picture; we have described these results
separately \citep{sio98}.  Moreover, we serendipitously detected two episodes
of transient metallic  absorption-line features---caused by clumpy material in
the stellar  wind from the \kd\ passing across the line of sight to the
WD---and have likewise described this phenomenon elsewhere (Bond et al.\ 2001). 

\section{{\HST\/} Observations}

Our \HST\/ observations consisted of twelve GHRS G160M spectra, centered at
Lyman-$\alpha$ (1216~\AA)\null. The aim of the program was to obtain spectra at
orbital phases 0.25 and 0.75, for optimum determination of the velocity
amplitude of the WD's spectroscopic orbit. In spite of the demanding spacecraft
scheduling requirements, all but two of our observations were successfully
obtained at mid-exposure times within $\pm$0.065 of the desired phase; however,
due to a misinterpretation of the scheduling software, our first two
observations were actually made one spacecraft Earth occultation interval later
than desired. Four of our spectra were obtained in 1993 with the pre-repair
GHRS, and the remaining eight with the COSTAR-corrected GHRS in 1994--1995. 
The latter group of spectra had, of course, significantly higher
signal-to-noise, and form the bulk of the useful data for our analysis.  

Each of the spectra had a dispersion of 0.07~\AA~diode$^{-1}$ and  covered an
interval of 36~\AA, which spans only the central 50\% of the Stark-broadened
\Lya\ absorption profile.  Half of the spectra were taken centered on the rest
wavelength of \Lya, and half were shifted by 1~\AA\ in order to lessen the
effects of  fixed-pattern features in the detector. Each observation was 29.0
(pre-COSTAR) or 36.3~min (post-COSTAR) long, and was broken up into 16
equal-length subexposures in order to look for spectroscopic variations on the
9.25-min rotational period of the WD\null. (Rotational variations were indeed
found, as discussed separately by Sion et al.\ 1998.)

The spectra were processed and calibrated using the standard STSDAS/GHRS
pipeline software.  The best available calibration files and appropriate
versions (for the pre- and post-COSTAR  observations) of CALHRS were used. 
Spectra of the Pt--Ne calibration lamp (WAVECAL observations) were taken prior
to, or immediately following,  each observation so that we could obtain the
best possible wavelength scales. The output spectra were all shifted during the
processing to heliocentric velocities.

Table~\ref{spec_tab} lists details of the twelve spectra used in our study. In
order to determine orbital phases, we noted that the times of primary eclipses
tabulated by \Ibanoglu\ et al.\ (1994) and Guinan \& Ribas (2001) indicate that
the orbital period was nearly constant throughout the mid-1990's. The following
ephemeris, which fits the nine published eclipse timings from 1992.0 through
1999.2 with a maximum error of 13~s, 
\begin{equation} 
T_0 = t_{\rm mid-eclipse} = {\rm HJD \,\, 2,440,610.05693} + 0.52118373 \,E \, , 
\end{equation} 
was used to calculate the phases listed in Table~\ref{spec_tab}.

Also given in Table~\ref{spec_tab} are the measured radial velocities,
which  will be discussed below.  

\begin{figure}
\begin{center}
\includegraphics[width=4.5in]{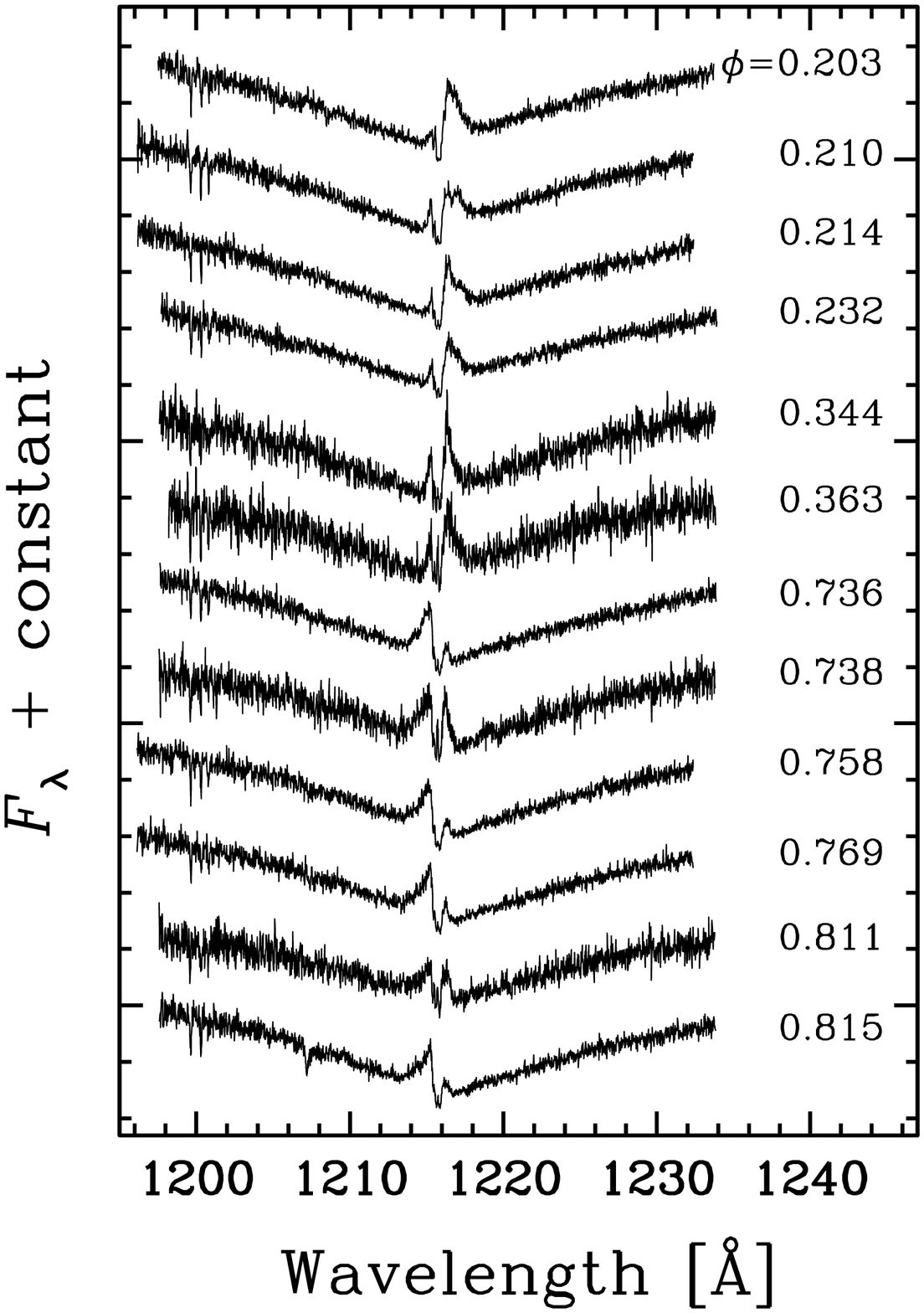}
\end{center}
\figcaption[f1.eps]{Our twelve GHRS Ly$\alpha$  spectra of
V471~Tau, shifted vertically for clarity. The spectra are presented in order of
orbital phase, labelled on the right.  The 4 pre-COSTAR spectra obtained in
1993 have noticeably lower S/N\null. Note the orbital motion of the broad
Ly$\alpha$ wings of the white dwarf, the out-of-phase motion of the
chromospheric emission core from the K~dwarf, and the stationary interstellar
absorption core. The bottom spectrum shows a transient absorption line of
\ion{Si}{3}, due to a mass ejection from the dK star passing in front of the
white dwarf (Bond et al.\ 2001).
\label{allspec}}
\end{figure}

\begin{figure}
\begin{center}
\includegraphics[angle=-90, width=\hsize]{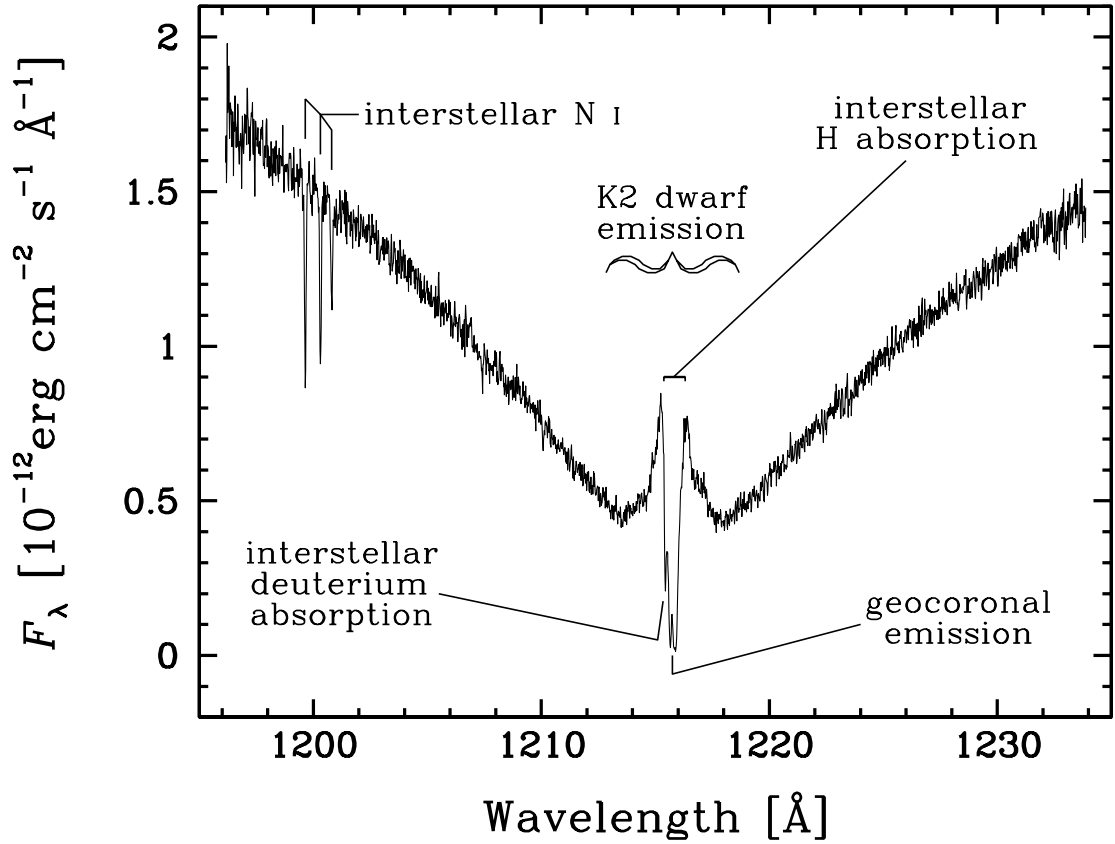}
\end{center}
\figcaption[f2.eps]{\Lya\ spectrum of V471~Tau, obtained by summing the
eight post-COSTAR spectra without any velocity shifts. The various features
seen superimposed on the broad absorption wings of the white dwarf are
labelled.
\label{coadd_unshift}}
\end{figure}

Figure~\ref{allspec} displays plots of all of the spectra obtained in our
study, in order of orbital phase. The lower S/N ratios of the four pre-COSTAR
observations are obvious. The spectra show a number of features, which are
explained in Figure~\ref{coadd_unshift}; this figure plots a coadded spectrum
made by summing the eight post-COSTAR observations, without any velocity
shifts.  Most obvious are the WD's broad and deep \Lya\ absorption wings. Three
sharp interstellar \ion{N}{1} absorption lines (1199.55, 1200.22, and
1200.71~\AA) lie near the short-wavelength end of the spectrum. The complicated
structure in the core of \Lya\ consists of  a broad chromospheric \Lya\
emission line from the K~dwarf, into which cuts a strong interstellar \Lya\
absorption core. The emission profiles are highly variable in structure and
strength, indicating varying levels of chromospheric activity and its
distribution across the stellar surface. An additional weak interstellar 
deuterium (\ion{D}{1} 1215.35 \AA) absorption is present on the blue side. 
Finally, there is a weak geocoronal \Lya\ emission feature in the center of the
strong interstellar absorption core.  On two occasions a transient absorption
line of \ion{Si}{3} 1206.51~\AA\ appeared suddenly in the spectra, which we
attribute to coronal mass ejections from the K~dwarf passing in front of the
WD; these are described separately (Bond et al.\ 2001).

Referring back to Figure~\ref{allspec}, one can clearly see the orbital motion
of the WD \Lya\ wings (blue-shifted in the top six spectra, red-shifted in the
bottom six), the out-of-phase motion of the K~dwarf's chromospheric emission
core, and the stationary interstellar and geocoronal features.  The bottom
spectrum, at orbital phase 0.815, shows a strong transient \ion{Si}{3}
1206~\AA\ absorption feature, as mentioned above.

\section{Spectroscopic Measurements}

\subsection{Radial Velocities}

Radial velocities of the WD were determined through a cross-correlation
technique.  We initially cross-correlated the 12 individual spectra with a
reasonable  model spectrum, which we calculated for $\Teff = 35,000$~K, $\logg
= 8.2$ using the TLUSTY/SYNSPEC software package  \citep{hub88,hub95}. Using
the  Levenberg-Marquardt least-squares algorithm \citep{pre92}, we determined
the velocity shift and flux  scale factor that best fit the model spectrum to
each observation (the flux scaling factor being necessary both to convert from
model fluxes  to observed fluxes, and because slightly different centerings of
the star in the GHRS aperture for the different observations led to small
changes in the total flux). We then used the resulting initial velocity
estimates to shift each individual spectrum to zero velocity, and summed all of
the spectra to obtain a grand ``zero-velocity''  spectrum with a much higher
S/N ratio than any of the individual spectra. After smoothing the grand
spectrum using a 16-point running average, we fitted it again to each
individual spectrum to obtain improved velocity estimates.  The process was
then iterated a second time in order to derive the final velocity measurements
(convergence having been achieved with the second iteration).

At each stage of the above process we  were careful to mask out the central
portions of each spectrum, which  contained the features from the K star and
interstellar medium described above, along with the region around the
interstellar  \ion{N}{1} triplet and the transient \ion{Si}{3} line.  Because
of suspected  differences in the absolute flux calibrations between the pre-
and  post-COSTAR data (see below), we actually created two different grand
spectra:  one each for the pre- and post-COSTAR data. We then cross-correlated 
the 4 individual pre-COSTAR spectra with the pre-COSTAR grand spectrum,  and
the 8 post-COSTAR spectra with the post-COSTAR grand spectrum, to  get the
final velocities.  Table~\ref{spec_tab} lists the resulting  velocities for all
12 spectra.  The listed errors are $1\sigma$, calculated from the least-squares
fit, and are of course larger for the lower-S/N pre-COSTAR spectra. The zero
points of the velocity scales are discussed in detail below (\S4).

\subsection{Effective Temperature and Gravity of the White Dwarf}

\begin{figure}
\begin{center}
\includegraphics[angle=-90, width=\hsize]{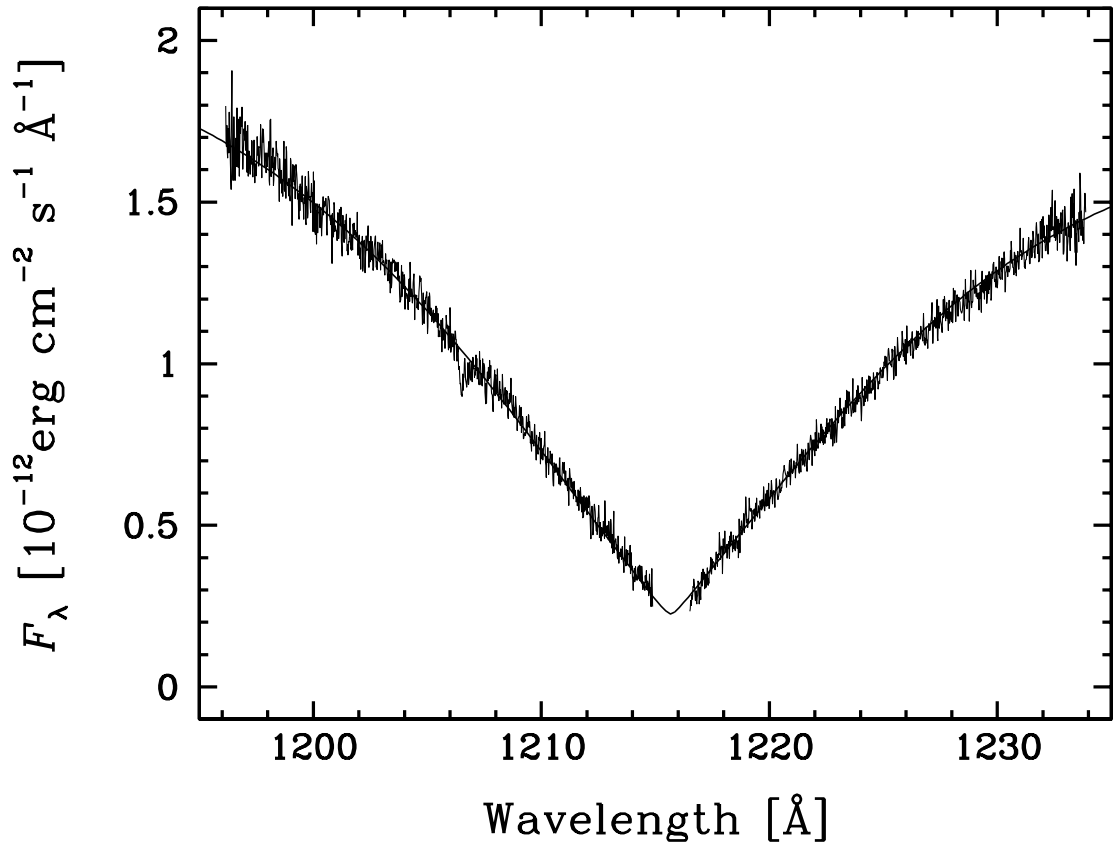}
\end{center}
\figcaption[f3.eps]{The \VTau\ \Lya\ profile, obtained by coaddition of the
eight post-COSTAR spectra, after shifting the individual spectra to zero
velocity. Superimposed as a noise-free line is a model-atmosphere profile
calculated for $\Teff=34,500$~K and $\log g=8.3$. Note the excellence of the
fit. A weak \ion{Si}{3} 1206~\AA\ line is present in the coadded sum, due to a
combination of a photospheric feature (Sion et al.\ 1998) and the transient
line due to silhouetted mass ejections from the K~star. \label{coadd_shift}}
\end{figure}

The final post-COSTAR coadded spectrum, based on a total of nearly 5 hours of
GHRS exposure time, gives us the opportunity to make a precise new
spectroscopic determination of $\Teff$ and $\logg$ for the WD component.  We
constructed a grid of pure-hydrogen TLUSTY models  covering the range $7.5 <
\logg < 8.7$ (in steps of 0.1~dex)  and $30,000 \, {\rm K}  < \Teff < 40,000
\,\rm K$ (in steps of 500~K), and then cross-correlated the  resulting
synthetic spectra (created using SYNSPEC) with the grand  coadded spectrum.  We
initially constructed both LTE and non-LTE models,  but the primary difference
between them is a sharper, deeper \Lya\ absorption core in the  NLTE models. 
Since this core is obliterated in our spectra by the K-dwarf and interstellar
features anyway, the LTE and NLTE models fitted the final  spectrum equally
well.  The best fit is obtained for $\Teff = 34,500 \pm 1,000$~K and $\logg =
8.3  \pm 0.3$---very close to the ``reasonable'' model used initially to 
estimate the velocities.  Figure~\ref{coadd_shift} shows the excellence of the
fit.  

Our results are in good agreement with previous determinations. In an early
study of the \Lya\ profile from \IUE\/ spectra, \citet{GuinanSion84} found
$\Teff = 35,000 \pm 3,000$~K and $\logg \simeq 8$. Cully et al.\ (1996) used
the EUV flux distribution to derive $\Teff = 33,100 \pm 500$~K (for an assumed 
$\logg = 8.5$), while from the same {\it EUVE\/} data Dupuis et al.\ (1997)
found 32,000~K\null. Barstow et al.\ (1997) and Werner \& Rauch (1997) have
used {\it ORFEUS\/} spectra of the Lyman series  to derive the effective
temperature and gravity. Barstow et al.\ obtain $\Teff=32,400^{+270}_{-800}$~K
and  $\log g=8.16^{+0.18}_{-0.24}$, while Werner \& Rauch find $35,125 \pm
1,275$~K and $8.21\pm0.23$, the primary difference in the analyses being the
inclusion of the \IUE\/ \Lya\ profile in the analysis by Barstow et al.

We will return to the temperature and gravity of the WD in \S7, in order to
discuss the star's radius and cooling age, and to constrain the evolutionary
history of the binary.

\section{The Double-Lined Spectroscopic Orbit}

We performed a least-squares fit of the final WD velocities to a sine
curve of the form
\begin{equation}
V(t) = \gamma - K_{\rm WD}\,\sin 
       \left[ \frac{ 2\pi(t - T_0)}{P_{\rm orb}} \right],
\end{equation}
where the systemic velocity, $\gamma$, and the WD's radial-velocity 
semi-amplitude, $K_{\rm WD}$, are the fitting parameters, and the time  of
mid-eclipse, $T_0$, and  the orbital period, $P_{\rm orb}$, are known (our
equation~[2]).  (The orbital eccentricity is assumed to be zero, in agreement
both with observation [Bois, Lanning, \& Mochnacki 1988] and theoretical
expectation.) We fitted the pre- and post-COSTAR data  separately.

%
%

In our initial fits, we found that, although the the pre- and post-COSTAR data
yielded the same $K_{\rm WD}$ to within the errors, the systemic $\gamma$
velocities were significantly different, with the pre-COSTAR value being more
negative by about $60\kms$. We believe we have traced this problem to the fact
that the absolute GHRS instrumental sensitivity falls off sharply across the
\Lya\ line; this means that the reduction of the instrumental spectra to
absolute fluxes involves multiplication by a function that has a very steep
slope with respect to wavelength. Detailed examination of the GHRS detector
sensitivity as a function of wavelength, kindly provided by S.~Hulbert (2000,
private communication) of the GHRS team at the Space Telescope Science
Institute (STScI), showed moreover that the pre- and post-COSTAR calibration
curves adopted in the reduction pipeline differ significantly from each other. 
It is thus perhaps not surprising that our analysis of a very broad spectral
feature would yield different velocity systems for the two sets of spectra. 
(This effect would be expected to be much smaller on sharp features, such as
the three interstellar \ion{N}{1} lines. We investigated their behavior by
first measuring velocities in the individual 29- and 36-min spectra; this
confirmed self-consistent velocities, within the errors, among the individual
pre- and post-COSTAR spectra.  We then, as described above, prepared grand
summed spectra for the 4 pre-COSTAR and 8 post-COSTAR spectra.  This revealed a
small systematic shift between the two sets of spectra, in the same sense as
that obtained from \Lya, but much smaller. For the pre-COSTAR grand spectrum,
the mean \ion{N}{1} velocities for the 1199.55, 1200.44, and 1200.71~\AA\ lines
were $+15.2 \pm 4.0$, $+19.5 \pm 5.4$, and $+21.0 \pm 7.8 \kms$, respectively,
while for the higher-quality post-COSTAR spectra the velocities were $+23.7 \pm
1.8$, $+23.0 \pm 2.3$, and $+23.5 \pm 3.3 \kms$. The weighted mean difference
between the pre- and post-COSTAR velocities was thus only about $6\kms$.)

In view of these differences, we decided, in the remainder of the analysis, to
use only the eight post-COSTAR spectra.  Moreover, although all of the 8
velocities should be on the same system, we feel that the absolute velocity
zeropoint for \Lya\ is not trustworthy because of the steep sensitivity
function.  

\begin{figure}
\begin{center}
\includegraphics[angle=-90, width=\hsize]{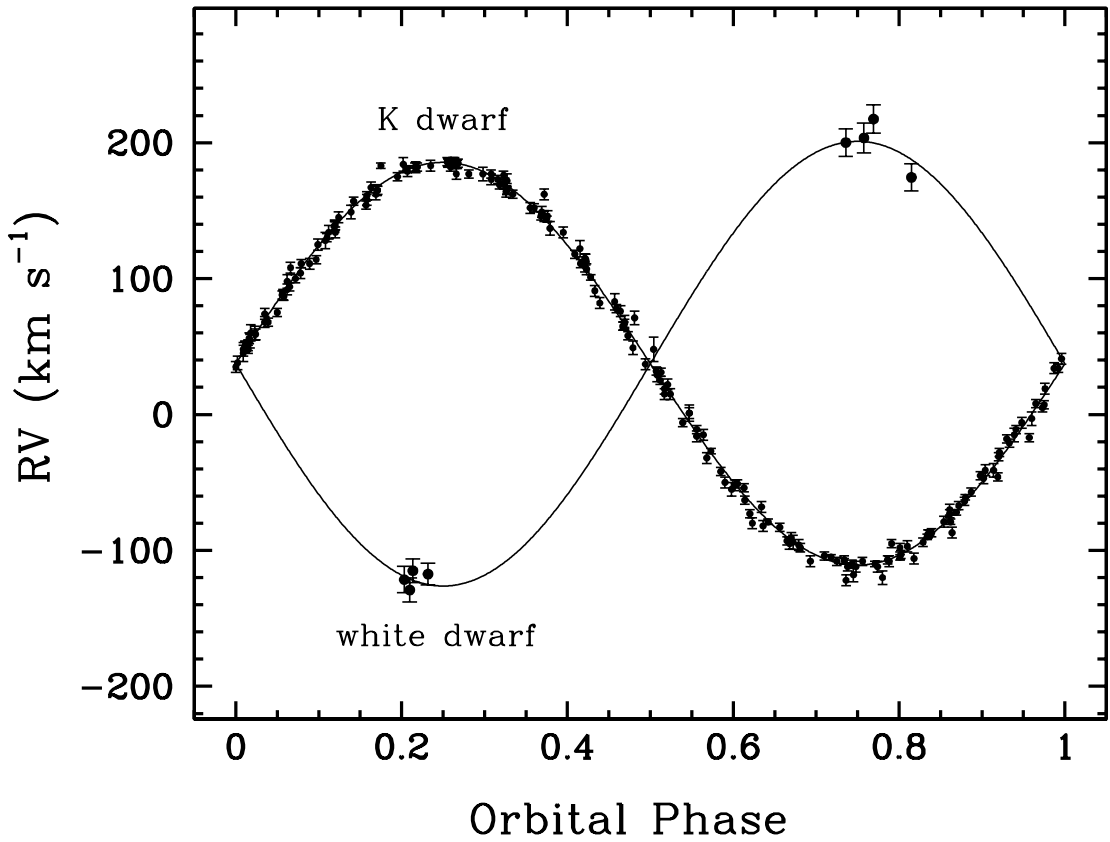}
\end{center}
\figcaption[f4.eps]{Radial velocity versus orbital phase for the
white-dwarf and K-dwarf components of V471 Tau.  Velocities for the white dwarf
are from the eight post-COSTAR \HST\/ observations discussed here, and the
numerous velocities for the K star are from the ground-based measurements of
Bois et al.\ (1988). Solid lines show the best-fitting sinusoidal fits for each
component. 
\label{rv}}
\end{figure}

Our final result for the WD velocity semi-amplitude is  $K_{\rm WD} = 163.6 \pm
3.5\kms$. For the dK star's semi-amplitude, we adopt $K_{\rm dK}=148.46 \pm
0.56 \kms$ from the ground-based spectroscopic orbit of Bois et al.\ (1988,
their Table~V; we took the mean of the two circular orbits presented in their
table, weighted by number of observations),

In Figure~\ref{rv} we plot the ground-based velocity observations of the
K~dwarf, taken from \citet{boi88}, our 8 post-COSTAR WD velocities, and the
best-fitting sine functions. For simplicity, we have plotted the WD velocity
curve with a $\gamma$ velocity assumed to be the same as that of the K~dwarf.
(Note that, unfortunately, the above discussion implies that we are unable to
make an independent measurement of the gravitational redshift of the WD based
on our \Lya\ observations.)  

\section{Component Masses}

The component masses and velocity semi-amplitudes are related through Kepler's
third law, which can be written as
\begin{equation}
M_{\rm (WD,\, dK)} \, \sin^3 i = 1.0385\times10^{-7} \,
 (K_{\rm WD} + K_{\rm dK})^2 \, K_{\rm (dK,\, WD)} \, 
 P_{\rm orb} \, ,
\end{equation}
where the respective masses of the two stars, $M_{\rm (WD,\, dK)}$, are in
$M_\sun$, the respective velocity semi-amplitudes, $K_{\rm (dK,\,
WD)}$, are in $\!\kms$, and the orbital period, $P_{\rm orb}$, is in
days.

In order to proceed, we need to determine, or place limits on, the inclination,
$i$, of the plane of the binary orbit to the line of sight. 

\subsection{Constraining the Orbital Inclination}

\begin{figure}
\begin{center}
\includegraphics[width=6.25in]{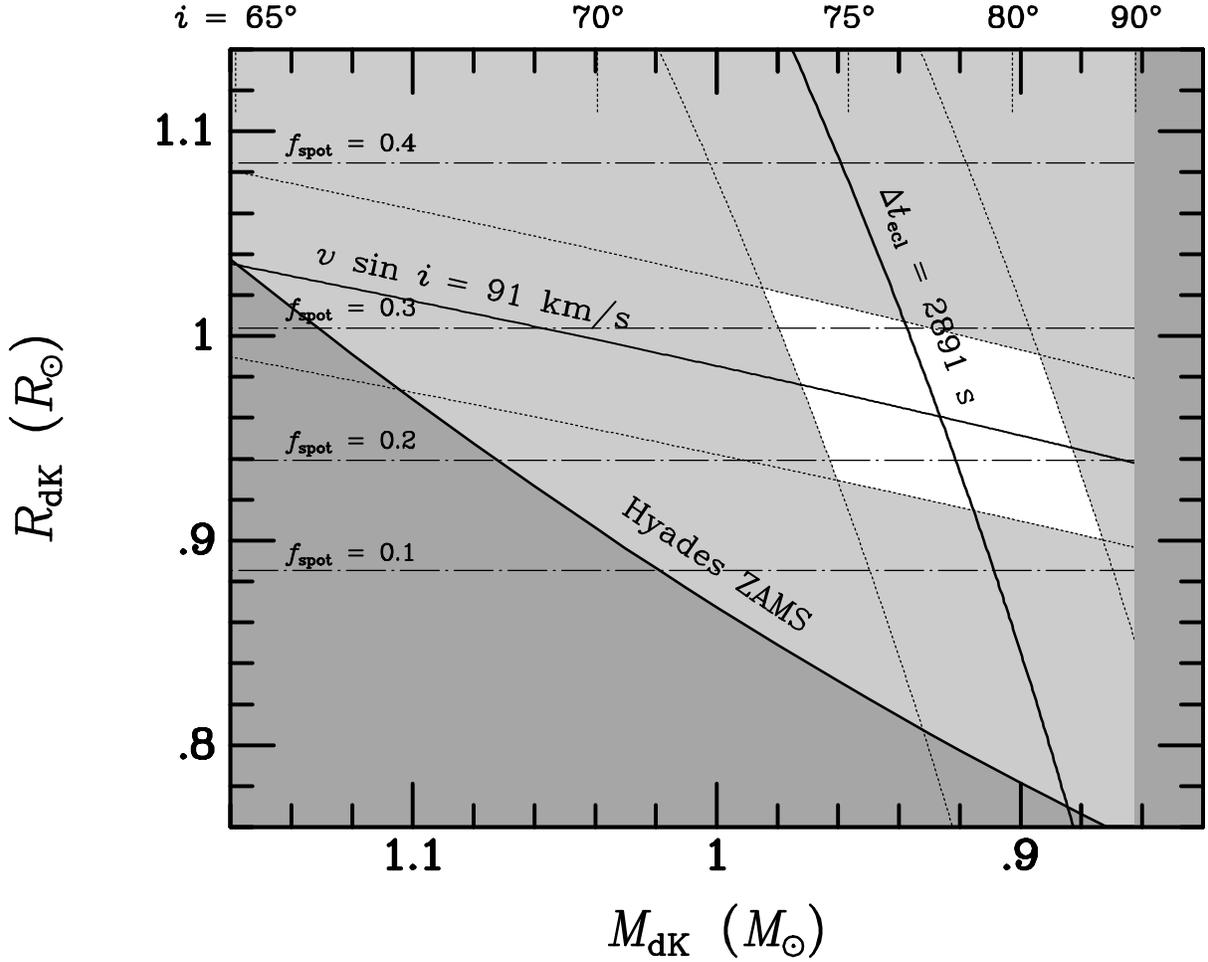}
\end{center}
\figcaption[f5.eps]{The mass-radius plane for the K-dwarf component of
V471~Tau. Values of the orbital inclination $i$ corresponding to the mass
coordinate are labelled at the top. The curve marked ``Hyades ZAMS'' shows the
main-sequence mass-radius relation for Hyades dwarfs (see text).  Dark shading
marks the excluded regions on the right-hand side corresponding to $i>90^\circ$
and at the lower left  below the Hyades ZAMS curve. Horizontal lines indicate
the radius of the star based on radiometry and the assumption that the
indicated fraction $f_{\rm spot}$ of the surface area is covered with dark
starspots. The nearly vertical curve labelled ``$\Delta t_{\rm ecl}=2891$~s''
(and the associated error limits and lightly shaded areas) delimit the allowed
values based on measurements of the duration of the total eclipse. The more
nearly horizontal curve marked ``$v \sin i=91$~km/s'' (and its associated error
limits and light shading) show allowed values based on the measured rotational
velocity. The intersection of the two curves shows our final solution for the
mass, radius, and orbital inclination, and the small white parallelogram shows
the error limits.
\label{massradius}}
\end{figure}

It will be useful in the following discussion to consider the location of the
K~dwarf in the mass-radius plane, which is shown in Figure~\ref{massradius}. We
immediately have a lower limit on $M_{\rm dK}$ by setting $i=90^\circ$ in
equation~(4) and adopting the values of $K_{\rm WD}=163.6 \pm 3.5 \kms$ and
$K_{\rm dK}=148.46 \pm 0.56 \kms$ from \S 4, and $P_{\rm orb} =
0.52118373$~day. This gives $M_{\rm dK} \ge 0.86 \pm 0.04 M_\sun$. The  region
excluded in this manner is shown in  Figure~\ref{massradius} as the vertical
shaded area on the right-hand side whose edge is labelled at the top with
$90^\circ$. Also shown as vertical ticks along the top edge are the values of
$M_{\rm dK}$ corresponding to $i=65^\circ$, $70^\circ$, $75^\circ$, and
$80^\circ$.

We can further limit the permitted region in Figure~\ref{massradius} by
assuming that the K~dwarf has a radius no smaller than that of a Hyades
zero-age main-sequence (ZAMS) star of the same mass. (Since the star is
constrained to rotate synchronously with the short-period orbit [see below], is
magnetically active, covered with starspots, and has relatively recently
emerged from a CE event, one cannot be sure that it is not {\it larger\/} than
a ZAMS star, but there seems to be no plausible way for it to be
smaller\footnote{Exceptions to this statement can occur under conditions of
high mass loss or mass accretion---e.g., Sarna, Marks, \& Smith 1995 and
references therein.}.) Perryman et al.\ (1998, their Table~10) have presented 
tables of computed values of the Hyades ZAMS masses, luminosities, and
effective temperatures, which agree well with observations. The ZAMS radii may
of course be calculated from these tables. The following formula fits the
Perryman et al.\ Hyades ZAMS mass-radius relation to  0.6\% accuracy or better
over the interval $0.8 \le M/M_\sun \le 1.2$:
\begin{equation}
 R/R_\sun = 0.7126 - 0.6277 (M/M_\sun) + 0.7826 (M/M_\sun)^2 \, .
\end{equation}
This relation, and the excluded triangular region below it, are shown by the
dark shaded area on the lower left-hand side of Figure~\ref{massradius}.

%
%
%
%

The K~star's radius is limited at the upper end by the fact that the binary
is detached, i.e., the K~star does not fill its Roche lobe. This limit is,
however, off the top of Figure~\ref{massradius}, at $R_{\rm dK} \approx
1.3\,R_\sun$ (see \S 5.2 below).

Further constraints on the radius of the K~dwarf come from radiometric
considerations.  \Ibanoglu\ et al.\ (1994), on the basis of their heroic
multi-decade photometric monitoring program on V471~Tau, report that the mean
brightness (outside eclipse) of the system increased almost monotonically by
0.21~mag over the interval from 1970 to late 1992, without an appreciable
change in $B-V$ color. This indicates strongly that a significant, and
time-variable, fraction of the K~star's surface is covered by dark spots, as is
indeed confirmed directly by the spectroscopic Doppler imaging of Ramseyer,
Hatzes, \& Jablonski (1995).  Hence a straightforward radiometric analysis will
yield only a lower limit to the physical radius of the star. The highest limit
will be obtained by taking the brightest $V$ magnitude that has ever been
observed, this corresponding to the smallest spot covering fraction of the area
of the visible hemisphere, $f_{\rm spot}$.  (In our simplified approach, 
$f_{\rm spot}$ refers to the equivalent fractional area of the hemisphere that
would be covered if the starspots were completely black.)

The brightest mean $V$ and $B$ magnitudes tabulated by \Ibanoglu\  et al.\ are
9.39 and 10.26, respectively (at epoch 1992.88), with errors of about
$\pm0.03$~mag. V471~Tau exhibits a photometric wave on its orbital period,
which arises from the patchy distribution of starspots across its surface. In
late 1992 the peak-to-peak amplitude of the rotational light curve was
$\sim$0.1~mag (Tunca et al.\ 1993), so we adopt $V_{\rm max} = 9.34$ and
$B_{\rm max} = 10.21$ as the very brightest magnitudes that it has reached. 
The WD has $V=13.64$ according to Barstow et al.\ (1997), who extrapolated
\IUE\/ spectra out to visual wavelengths; the result agrees well with the
$V=13.6$ quoted by Warner, Robinson, \& Nather (1971) from direct measurements
of the eclipse depth in the $V$ band. The color of the WD is $B-V=-0.24$ (from
the model atmospheres of Bergeron et al.\ 1995). The WD therefore contributes
2\% of the system light at $V$ and 6\% at $B$, leading to corrected values of
$V_{\rm max} = 9.36$, $B_{\rm max}=10.27$, and $(B-V)_{\rm max}= 0.91\pm0.04$.
(At our level of approximation, we are neglecting the facts, discussed by
Ramseyer et al.\ 1995 in more detail, that a $\sim$3\% ellipsoidal orbital
variation arises from tidal distortion, and that there is a small amount
[$\sim$100~K] of heating of the dK star on its sub-WD hemisphere.) Note that
this color is in quite good agreement with the value of $B-V=0.92$ actually
measured during primary eclipse (thus isolating the K~dwarf) by \citet{you72},
but that the K~star had brightened appreciably in 1992 from the $V=9.71$ that
Young \& Nelson measured during primary eclipse in the early 1970's. 

The direct trigonometric parallax of V471~Tau from the {\it Hipparcos\/}
mission is $21.37\pm1.62$~mas.  More recently, however, de~Bruijne et al.\
(2001) have used {\it Hipparcos-Tycho\/} proper motions to determine a more
accurate secular parallax of $21.00\pm0.40$~mas (corresponding to
$d=47.6\pm0.9$~pc), based on the constraint that the star shares the space
motion of the Hyades cluster. This distance then implies a brightest absolute
magnitude  $M_{V,\rm\,max} = 5.97 \pm 0.06$ for the dK
component\footnote{\citet{per98} use $B-V = 0.83$ to place \VTau\ (which they
catalog as HIP~17962) about 0.2~mag below the Hyades ZAMS\null. This however is
the color of the system {\it outside\/} eclipse, which, as noted, includes some
contamination from the WD component. Our values place the star on the Hyades MS
in the $M_V, B-V$ diagram, although this is somewhat fortuitous given the
star's large amount of spot coverage (see \S6 below).}.

Using the photometric model library of \citet{lej97}, and assuming
$\rm[Fe/H]=0.1$ and no interstellar reddening, we find that a color of
$B-V=0.91\pm0.04$ implies that the \kd\ has an effective temperature of  $\Teff
= 5040 \pm 100$~K\null. (More precisely, this is the temperature of the
unspotted portion of the stellar surface.) The corresponding bolometric
correction  is ${\rm BC}(V) = -0.28 \pm 0.02$, yielding $M_{\rm bol,\,max} =
5.69 \pm 0.06$ and (adopting $M_{\rm bol,\sun} = 4.746$ from Lejeune et al.\
1997), we  derive $L_{\rm dK,\, max}/ L_\sun= 0.42 \pm 0.02$.  The radius of
the K star is then
\begin{equation}
 R_{\rm dK}/R_\sun = (0.84 \pm 0.04) (1-f_{\rm spot})^{-1/2}  .
\end{equation}

%


Doppler maps of the K star by Ramseyer et al.\ (1995) suggest that $f_{\rm
spot}\approx0.2$ in late 1992; since this estimate is based only on our visual
inspection of their published maps, and the starspots that they have detected
may not be completely black, the uncertainty in $f_{\rm spot}$ is probably of
order a factor of two.  The derived values of $R_{\rm dK}/R_\sun$ for $f_{\rm
spot}=0.1$, 0.2, 0.3, and 0.4 are shown as four horizontal dot-dash lines in
Figure~\ref{massradius}, labelled with the corresponding values of $f_{\rm
spot}$. We believe that it is likely that the star's radius lies between the
top and bottom lines.

The next constraint on $R_{\rm dK}$ is a geometrical one that comes from the
duration of primary eclipse.  It can be shown (e.g., Young \& Nelson 1972,
their eq.~[1]) that the radius of the K~star is related, to good
approximation, to the orbital phase, $\phi$, at which the eclipse ends by
\begin{equation}
 R_{\rm dK} = a \, (\sin^2 i \, \sin^2 \phi + \cos^2 i)^{1/2} ,
\end{equation}
where $a$ is the semi-major axis of the orbit. We define $\phi = (0.5 \,\Delta
t_{\rm ecl}/P_{\rm orb}) \times 360^\circ$, where $\Delta t_{\rm ecl}$ is the
duration of the eclipse measured from mid-ingress to mid-egress.  The duration
is given by $\Delta t_{\rm ecl}=\Delta t_{12} + \Delta t_{23}$, where (in the
notation of Warner et al.\ 1971) $\Delta t_{12}$  is the time
interval between first and second contacts (i.e., the duration of eclipse
ingress) and $\Delta t_{23}$ is that between second and third contacts (i.e.,
the duration of totality). Based on the high-speed photometry of  Warner et
al., we adopt  $\Delta t_{12}=68.3 \pm 2.0$~s and $\Delta t_{23}=2822.6 \pm
2.0$~s, so that $\Delta t_{\rm ecl}=2890.9 \pm 2.8$~s and $\phi = 11\fdg56 \pm
0\fdg01$. 

The separation of the stars is given by
\begin{equation}
a\sin i = 1.3751\times10^4 \,(K_{\rm WD}+K_{\rm dK})\,P_{\rm orb} \quad\rm km,
\end{equation}
where the velocity semi-amplitudes are in $\!\kms$ and $P_{\rm orb}$ is in
days. Substituting our measurements of $K_{\rm WD}$ and $K_{\rm dK}$ into
equation~(8), and then substituting $a$ and $\phi$ into equation~(7), we obtain
a relation between $R_{\rm dK}$ and $\sin i$, which is plotted in
Figure~\ref{massradius} as the curve labelled ``$\Delta t_{\rm ecl}=2891$~s,''
along with the 1$\sigma$ error limits on either side. Regions of the plot
outside the allowed limits are lightly shaded.

Yet another constraint on the K~dwarf's radius can be derived from its observed
rotational velocity, under the assumption that it rotates synchronously with
the orbital period (i.e., $P_{\rm rot,\,dK} = P_{\rm orb}$). This assumption is
strongly supported by the fact that photometric variations of the system
arising from starspots on the dK component occur almost precisely on the
orbital period (e.g., \Ibanoglu\ 1978; Skillman \& Patterson 1988 and
references therein). On the further assumption that the rotational axis is
perpendicular to the orbital plane, the projected rotational velocity is given
by
\begin{equation}
v_{\rm rot,\,dK}\,\sin i = \frac{2 \pi R_{\rm dK}}{P_{\rm orb}} \sin i.
\end{equation}

The rotational velocity of the K~dwarf has been measured to be $v_{\rm
rot,\,dK}\,\sin i = 91 \pm 4\kms$ by Ramseyer et al.\ (1995). The family of
points satisfying equation~(9) is plotted in Figure~\ref{massradius} as the
curve labelled ``$v \sin i=91\rm\,km/s$,'' again with the associated 1$\sigma$
error limits plotted on either side. Again we have shaded the excluded regions
of the plot outside the allowed limits.

Figure~\ref{massradius} now shows that the mass and radius of the K~star have
been fairly tightly constrained to the small unshaded parallelogram-shaped box,
based on our measurement of the WD's velocity amplitude combined with purely
geometrical measurements.  The  orbital inclination corresponding to the
intersection of the two relations is
\begin{equation}
\sin i = 0.976 \pm 0.020 ,
\end{equation}
or $i = 77.4^{+7.5}_{-4.5}$~degrees. Note also that the radiometric radius
determination discussed above would be in good agreement with these
constraints, if the hemispheric spot-covering fraction in late 1992 during the
brightest phase of the rotational light curve was $f_{\rm spot} \approx
0.2$--0.3. The physical radius of the K~star corresponding to the intersection
of the two relations is $R_{\rm dK}=0.96\pm0.04 \, R_\sun$. 


\subsection{The Stellar Masses and Orbital Parameters}

We now have all of the ingredients required to calculate the masses directly 
from the orbital parameters.  According to equations~(4) and~(10), combined
with the measurements of $K_{\rm WD}$ and $K_{\rm dK}$, the WD and K-dwarf
masses are: 
\begin{eqnarray}
M_{\rm WD} &=& 0.84 \pm 0.05 \, M_\sun, \nonumber\\
M_{\rm dK} &=& 0.93 \pm 0.07 \, M_\sun. \nonumber
\end{eqnarray}

Most previous authors have followed Young \& Nelson (1972) in adopting masses
near 0.7--$0.8\,M_\sun$ for both components of V471~Tau. Until now, it was
necessary to {\it assume\/} that the dK star had a normal ZAMS mass and radius
appropriate to its spectral type, since the velocity amplitude of the WD
component was unknown, and then to derive other parameters of the system from
the single-lined spectroscopic orbit of the K~component.  Now that  the
velocity curves for both stars are available (along with the crucial
measurements of the eclipse timings and the K~star's rotational velocity), we
can derive the system parameters without recourse to assumptions about the K
star---and indeed we have found that assumption of a normal ZAMS mass and
radius for the K star would be incorrect. 

Other orbital parameters also follow from our measurements. The orbital
separation of the two stars (from equation~[8]) is $a=3.30\pm0.08\, R_\sun =
3.43\pm0.08\, R_{\rm dK}$\null. The Roche-lobe radii (from equation~[18], given
below in \S 8) of the two stars are $R_{L,\rm WD} = 1.22 \pm 0.10 \, R_\sun$ and
$R_{L,\rm dK} = 1.28 \pm 0.09 \, R_\sun$.  The K~star fills its Roche lobe by
only $R_{\rm dK}/R_{L,\rm dK} = 73\% \pm 6\%$, so it is well inside its lobe.

We will now discuss some of the physical properties of each component star in
more detail.

\section{The Oversized K Star}

Figure~\ref{massradius} illustrates that the dK star in V471~Tau is about 18\%
oversized for its mass, relative to normal Hyades ZAMS stars. This conclusion
has already been suggested by previous authors (e.g., Ramseyer et al.\ 1995),
but is strengthened by our new analysis.  

What about the location of the K~dwarf in the mass-luminosity plane? A normal
Hyades ZAMS star of $0.93\,M_\sun$ would have $L/L_\sun = 0.40$ (Perryman et
al.\ 1998, their Table~10). To estimate the actual luminosity of the dK star in
V471~Tau, it is appropriate to use the mean system brightness (thus averaging
over the entire surface of the spotted star). In late 1992, the mean brightness
was $V=9.41$ (\Ibanoglu\ et al.\ 1994, corrected for the WD contribution; cf.\
\S5.1). Using the bolometric correction and solar $M_{\rm bol}$ from \S5.1 and
the known distance, we find $L/L_\sun = 0.40$ for the dK star, in perfect
agreement with the expected luminosity for the star's mass.  This can be
compared with the fact that, for the mass of $\approx$0.7--$0.8\,M_\sun$
assumed in earlier literature, a Hyades ZAMS star would have $L/L_\sun \le
0.20$, according to Perryman et al.  (Our result, however, applies
strictly only to the 1992 epoch, and we would have gotten a somewhat different
result at earlier epochs when the star was somewhat fainter.)

We conclude that the K~star has a luminosity that is approximately appropriate
for its measured mass, but a radius that is larger than that of a ZAMS star of
the same mass. We will discuss two possible reasons for the unusually large
radius of the K~star: (a)~the star is still out of thermal equilibrium because
of its recent immersion in a CE; or (b)~the star is so covered by starspots
that convective energy transport is inhibited, causing the radius to increase.
(Irradiation of the dK star by the WD companion appears to be a minor effect,
which we will not consider further; the K star intercepts about 2\% of the
luminosity of the WD, or only about 1\% of its own luminosity. Tidal heating is
also negligible, given the synchronous rotation and circular orbit.)


The response of MS stars to CE events has been studied theoretically
by Hjellming \& Taam (1991). Although the star does go out of thermal
equilibrium while it is immersed in the CE, it is expected to revert very
quickly to its original configuration once the envelope has been ejected. This
reversion occurs on the thermal  timescale of the accreted material, which is 
$\sim$$10^3$--$10^4$~yr according to  Hjellming \& Taam.  Since V471~Tau
emerged from its CE about $10^7$~yr ago (see \S7.3 below), it
should have relaxed back to its ZAMS radius long ago. Thus the CE interaction
seems not to be the cause for the current oversized radius.


The second hypothesis appears to have more merit. As discussed above, the
surface of the active dK star is partially covered by starspots, which blocked
$\approx$25\% of the surface area of the star in the early 1990's. The response
of MS stars to starspot covering has been considered by Spruit (1982) and
modelled theoretically by Spruit \& Weiss (1986), who show that the response
depends on the timescale with which the spot covering fraction varies. If the
covering fraction were to increase, say, on a timescale short compared to the
thermal timescale of the stellar envelope\footnote{Sometimes loosely called the
``Kelvin-Helmholtz timescale of the envelope,'' this is $\approx$$10^5$~yr for
the Sun (Spruit 1982).}, the blocked luminosity would simply go into increasing
the energy content of the envelope, and the stellar radius and the surface
temperature outside the spots would remain essentially unchanged. The
photometric behavior of V471~Tau over the past few decades, described above,
agrees with this prediction---there is a change in overall brightness with
little or no change in color. However, if the star is permanently covered with
spots, over timescales long compared to the thermal timescale, Spruit \& Weiss
argue that both the surface temperature and radius will increase relative to
their values for a spot-free star. This is because, as described by Spruit \&
Weiss (1986), the luminosity produced by the core is affected only slightly
(for stars of $\approx$$1M_\sun$, which have only shallow convective layers)
and must continue, time-averaged, to escape from the star.

Our observation is thus in rough qualitative agreement with the theoretical
expectation. However, quantitatively, Spruit \& Weiss predict that, for stars
of $\approx$0.9--1.0~$M_\sun$, the stellar radius and surface temperature will
both increase by nearly the same factors of about 4\% each (for a spot covering
fraction of 25\%), whereas we actually find an increase of $\sim$18\% in radius
and a small {\it decrease\/} in the surface temperature. (Part of the
difference may be that Spruit \& Weiss calculate $T_{\rm s}$, the ``surface
temperature'' of the spot-free regions of the photosphere, while the
temperature that we infer from the observed $B-V$ color is actually an average
of the hotter spotless regions and the cooler, but not completely dark, spot
umbrae and penumbrae.)


We can estimate what the time-averaged effective spot covering fraction, $f$,
has to be in order for V471~Tau to maintain the luminosity of a $0.93\,M_\sun$
ZAMS star, as follows:
\begin{equation}
R_{\rm ZAMS}^2 \, T_{\rm eff,\,ZAMS}^4 = 
  (1-f)R_{\rm dK}^2\, T_{\rm surf,\,dK}^4 \, , 
\end{equation}
where $T_{\rm surf,\,dK}$ is the temperature of the unspotted portion of the
surface, and $f$ is averaged over the entire surface of the star, whereas in
\S5.1 we defined $f_{\rm spot}$ to be the minimum coverage fraction for the
visible hemisphere seen during the 12.5-hr rotation period. From Perryman et
al.\ (1998) we have $R_{\rm ZAMS}=0.806\,R_\sun$ and  $T_{\rm
eff,\,ZAMS}=5105$~K, and for the V471~Tau K~star, we found $R_{\rm
dK}=0.96\,R_\sun$ and $T_{\rm surf,\,dK}=5040$~K, from which we deduce
$f\simeq0.26$, in quite good agreement with the observations. It thus appears
that, on average, the spot covering fraction has been close to its current
value for at least the past $\approx$$10^5$~yr.

Note also that Mullan \& MacDonald (2001) have presented evidence that
magnetically active, but single, M dwarfs have significantly larger radii than
inactive dM stars of the same effective temperature. This may well be another
manifestation of the phenomenon seen in V471~Tau.  The cool component in the
pre-cataclysmic binary BE~UMa was found by Ferguson et al.\ (1999) to be larger
by about a factor of 2 than a MS star of the same mass; however, since BE~UMa
is surrounded by a faint planetary nebula, it must have emerged from its CE
very recently, and the large radius may be due to the cool star still being out
of thermal equilibrium.


\section{The Paradoxical White Dwarf}

We now turn to the other component of the V471~Tau binary, the $0.84\,M_\sun$
WD\null. We will first show that it is by all appearances a normal, hot white
dwarf---which in fact provides the most stringent test yet of the theoretical
mass-radius relation for WDs---and then show that this normality actually
raises a serious paradox involving the evolution of the binary system.

\subsection{Radius of the White Dwarf}

We have two methods available to determine the radius of the WD: indirectly
from radiometry, and directly from eclipse ingress timings.

The radiometric method is based on knowledge of the effective temperature and
distance of the star, combined with a measurement of its absolute flux.
Determinations of $R_{\rm WD}$ have been made by \citet{bar97} and
\citet{WernerRauch97}, based on the {\it Hipparcos\/} trigonometric parallax of
$21.37\pm1.62$~mas, effective temperatures of  $\Teff=32,400^{+270}_{-800}$ and
$35,125\pm1,275$~K, respectively (see \S3.2), and absolute fluxes from {\it
ORFEUS\/} and \IUE\null. The resulting radii reported by these authors were
$0.0107 \pm 0.0009 \, R_\sun$ and $0.0097 \pm 0.0013 \, R_\sun$, respectively,
the primary source of the difference being the different adopted $\Teff$
values.

We have recalculated the radius as follows: first we adopt our new temperature
determination (from \S3.2) of $\Teff=34,500 \pm 1,000$~K (which is close to the
mean of the two determinations above), and second we use the new, slightly
smaller and more precise secular parallax of  de~Bruijne et al.\ (2001),
$21.00\pm0.40$~mas. Applying these two adjustments to the Werner \& Rauch
calculation yields $R_{\rm WD} = 0.0102 \pm 0.0011 \, R_\sun$, with the
remaining error coming mostly from the $\sim$10\% uncertainty of the {\it
ORFEUS\/} absolute flux measured at 1100~\AA\null.  A nearly identical result
is obtained if the Barstow et al.\  radius is similarly adjusted to the new
temperature and distance.

The direct geometric method for determining the radius is based on the duration
of eclipse ingress (or egress), which we denoted $\Delta t_{12}$ in \S5.1.
Following the discussion in \S5.1, we adopt  $\Delta t_{12}=68.3 \pm 2.0$~s
from Warner et al.\ (1971). (Other authors, e.g.\ \Ibanoglu\ 1978 and
references therein; Beavers, Oesper, \& Pierce 1979; and Guinan \& Ribas 2001,
have given somewhat different values for $\Delta t_{12}$, ranging from 55 to
76~s, but the data of Warner et al., based on high-speed photon-counting
photometry with a 2-m class telescope, appear to have the highest quality for
this purpose.) 

The geometry of the eclipses in V471~Tau has been discussed by Warner et al.\
(1971) and Young \& Nelson (1972). The large difference in stellar radii,
$R_{\rm dK} \gg R_{\rm WD}$, means that the eclipse of the WD star by the limb
of the dK companion can be treated as the passage of a straight edge over the
WD\null. It can then be shown that the latitude on the dK star at which the WD
eclipses occur as seen from Earth, $\beta$, is related to the eclipse timings
by
\begin{equation}
\cos\beta = \left( \frac{R_{\rm WD}}{R_{\rm dK}} \,
  \frac{\Delta t_{\rm ecl}}{\Delta t_{12}} \right)^{1/2}
\end{equation}
(see Warner et al.\ 1971, their eq.~[5], and Young \& Nelson 1972, their
eq.~[2]).

The relative velocity of the two stars across the line of sight at mid-eclipse
is $V_{\rm dK}+V_{\rm WD} = (K_{\rm dK}+K_{\rm WD})/\sin i$, and the length of
the WD's path behind the dK star is approximately $2 R_{\rm dK} \cos\beta$. The
eclipse duration is thus given to good approximation by 
\begin{equation}
 \Delta t_{\rm ecl} = 2 R_{\rm dK} \cos\beta \sin i / 
   (K_{\rm WD}+K_{\rm dK}).
\end{equation}

By eliminating $\cos\beta$ from equations~(12) and~(13) and rearranging, we
find
\begin{equation}
 R_{\rm WD} = \frac{\Delta t_{12} \Delta t_{\rm ecl}(K_{\rm WD}+K_{\rm dK})^2} 
               {4 R_{\rm dK}\sin^2 i} \, .
\end{equation}

Inserting the measured values, we obtain $R_{\rm WD} =
0.0109\pm0.0007\,R_\sun$. The agreement with the radiometric radius is
remarkably good, especially considering that the observational determination of
$\Delta t_{12}$ is both technically challenging and somewhat dependent upon the
limb-darkening law of the WD, as discussed in detail by Warner et al.

\subsection{Location in the Mass-Radius Plane and Core Composition}

\begin{figure}
\begin{center}
\includegraphics[angle=-90, width=\hsize]{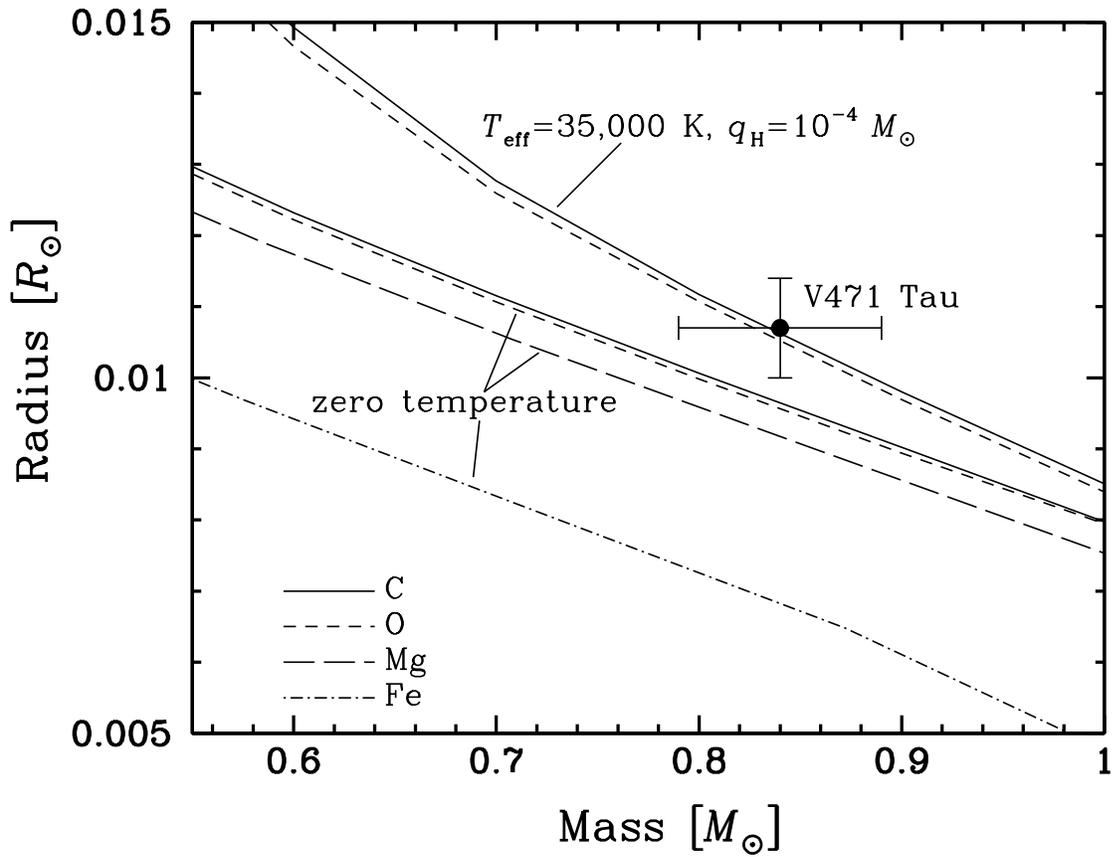}
\end{center}
\figcaption[f6.eps]{Position of the white-dwarf component of V471 Tau  in
the mass-radius plane.  For comparison we plot theoretical relations for C and
O white dwarfs of zero temperature and 35,000~K surface temperature (Wood
1995), and zero-temperature Mg and Fe white dwarfs (Hamada \& Salpeter 1961).
The agreement with the 35,000~K CO relation is excellent.
\label{mass_vs_rad}}
\end{figure}

Figure~\ref{mass_vs_rad} shows the location of the WD component of V471~Tau in
the mass-radius plane. We use our dynamical mass of $M_{\rm WD} = 0.84 \pm 0.05
\, M_\sun$ and a radius $R_{\rm WD} = 0.0107 \pm 0.0007 \, R_\sun$, which is
the weighted mean of the radiometric and geometric radii discussed in the
previous subsection.

For comparison with theory we plot mass-radius relations in
Figure~\ref{mass_vs_rad} from calculations of Wood (1995, and 1999 private
communication) for carbon and oxygen WDs with a surface hydrogen-layer mass of
$10^{-4} M_\sun$. We show Wood's relations both for zero-temperature WDs and
for WDs with an effective temperature of 35,000~K, very close to that measured
for the V471~Tau WD\null. In addition, we plot the zero-temperature relations
for WDs composed of Mg and Fe, from Hamada \& Salpeter (1961).

The figure shows that Mg and Fe core compositions are excluded. However the
agreement with C and O compositions is excellent, and indeed, at about the
1$\sigma$ level, we even confirm that the CO models with a 35,000~K surface
temperature provide a better fit than the zero-temperature models.  

Our observations provide perhaps the most stringent test and confirmation yet
made of the theoretical relation; it is rivaled in accuracy only by two or
three WDs with dynamical masses from visual-binary orbits and radiometric radii
based on trigonometric parallaxes (see Provencal et al.\ 1998). 

We conclude that the WD in V471 Tau matches the theoretical structure of a
normal CO WD with the star's measured mass, radius, and surface temperature.

\subsection{Location in the H-R Diagram and Cooling Age}

\begin{figure}
\begin{center}
\includegraphics[angle=-90, width=\hsize]{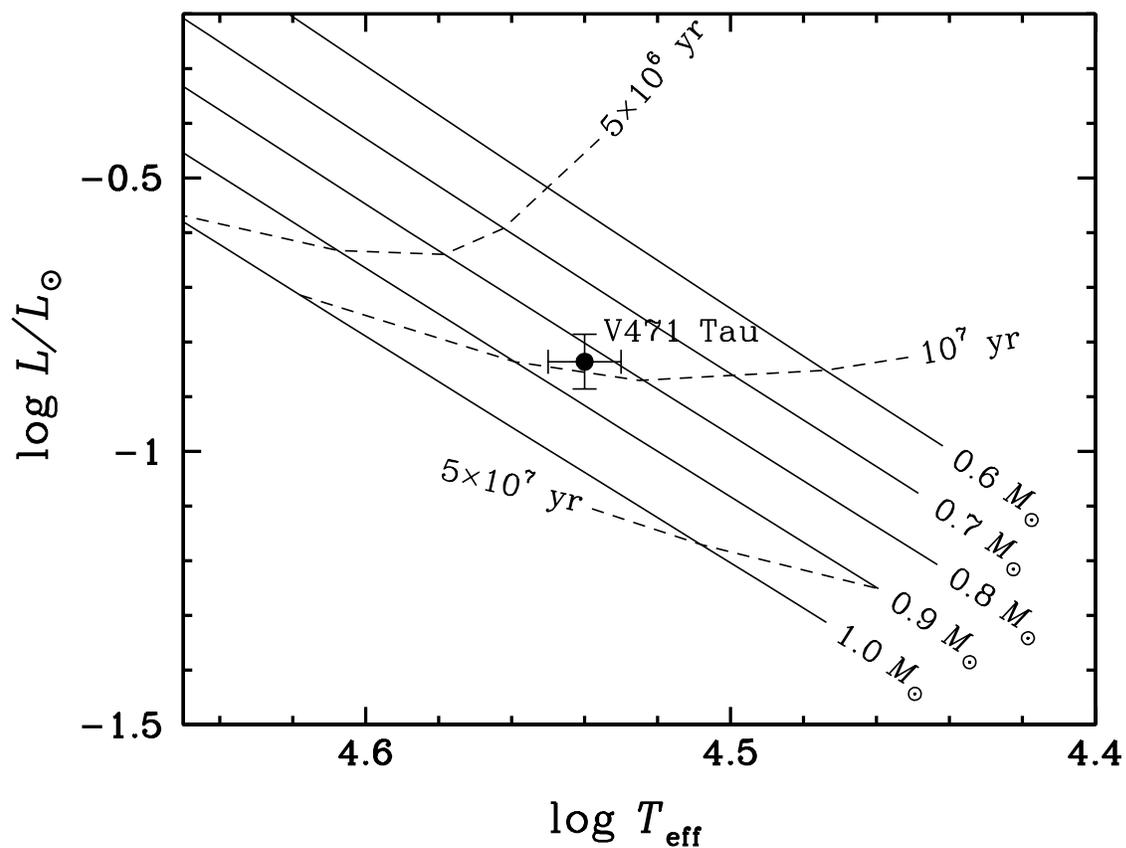}
\end{center}
\figcaption[f7.eps]{
The position of the V471~Tau white dwarf in the theoretical H-R diagram. Also
shown are cooling tracks for CO white dwarfs of various masses from Wood
(1995).  Dashed lines show the cooling ages. The position of V471 Tau is in
excellent agreement with that expected from its dynamically measured mass of
$0.84\,M_\sun$. Its cooling age is close to $10^7$~yr.
\label{hrd}}
\end{figure}

We can also plot the position of the WD component in the theoretical H-R
diagram, $\log\Teff$ vs.\ $\log L/L_\sun$\null. The effective temperature and
radius have been derived above, and the luminosity is given by $L_{\rm WD} =
(R_{\rm WD}/R_\sun)^2 \, (T_{\rm eff,\, WD}/T_{\rm eff,\,\sun})^4 \, L_\sun =
0.146 \pm 0.016 \, L_\sun$. The resulting position of the WD is shown in
Figure~\ref{hrd}. We also show the cooling tracks of Wood (1995, and 1999
private communication), along with the cooling ages, as functions of WD mass
from 0.6 to $1.0\,M_\sun$\null. As noted in the previous subsection, these
tracks are for CO white dwarfs with a surface hydrogen-layer mass of
$10^{-4}\,M_\sun$. 

The figure shows that the position of the WD is in excellent agreement with
that expected from its dynamically measured mass of $0.84\,M_\sun$\null. It
should be noted that the $\Teff$ was determined purely spectroscopically, and
that one of the two accordant methods used to determine the radius was based
only on spectroscopic and photometric methods. Thus the information used to
plot its position in Figure~\ref{hrd} is essentially independent of the star's
distance, and independent of its mass.  The fact that the WD lies almost
exactly on the theoretical position expected for its dynamically measured mass
is therefore a remarkable and gratifying confirmation of WD theory.  Indeed, if
we were to read off the mass from  Figure~\ref{hrd}, we would find
$0.83\pm0.06\,M_\sun$, in quite astonishing agreement with the dynamical
measurement.

As just noted, the luminosity of the WD can be calculated without any recourse
to distance measurements, if we use only the radius of
$0.0109\pm0.0007\,R_\sun$ that we determined from the eclipse ingress timings.
Hence we can combine the resulting luminosity with a brightness measurement to
determine a distance. The calculated luminosity based only on the
ingress-determined  radius would be $0.151\pm 0.026 \, L_\sun$\null. The
corresponding bolometric correction, from Bergeron et al.\ (1995), is ${\rm
BC}(V)=-3.30$, so the calculated absolute magnitude of the WD is $M_V=+10.10$.
Adopting the observed $V$ magnitude of 13.6 (Warner et al.\ 1971), we find a
distance modulus $m-M=3.50\pm 0.20$, corresponding to a distance of $50.1\pm
4.8$~pc and a parallax of $20.0\pm 1.8$~mas. This agrees very well with the
measured secular parallax of $21.00\pm0.40$~mas.

The cooling age of the WD is seen from Figure~\ref{hrd} to be very close to
$10^7$~yr.

For convenience we summarize all of the measured and inferred parameters of the
V471~Tau system in Table~\ref{param_tab}.

\subsection{Dynamical Validation of the Spectroscopic Gravity}

The dynamical mass and radiometric\slash geometric radius imply a WD surface
gravity of $\log g = 8.31 \pm 0.06$, in excellent agreement with the value of
$8.3 \pm 0.3$ that we derived from fitting model-atmosphere spectra to the
\Lya\ line, as discussed in \S3.2. This is an extremely reassuring validation
of the method of spectroscopic determination of surface gravity via
model-atmosphere line-profile fitting.

\subsection{The Paradox of the Temperature and Mass}

The WD in V471~Tau is too hot for its mass. This is clearly indicated by
Table~\ref{hyad_tab}, which presents the masses and temperatures of WDs in the 
Hyades cluster. Except for V471~Tau itself, the data are taken from the
literature, as indicated in the notes to the table.  Masses are available from
three different techniques: (1)~spectroscopic binary orbits, which yield
dynamical masses,  are available for V471~Tau (this paper) and for HZ~9 (which
does not eclipse, making it necessary to assume a mass for the cool companion
from its spectral type); (2)~gravitational redshifts (where the center-of-mass
velocity is known from the assumption that the WD shares the motion of the
Hyades cluster), and (3)~spectroscopic temperatures and surface gravities,
which yield a mass when combined with a theoretical mass-radius relation. The
columns in Table~\ref{hyad_tab} contain the following information: (1)-(2)~the
star name and Villanova WD catalog number; (3)~the dynamical mass; (4)~the mass
derived from the gravitational redshift; (5)~the mass derived from the
spectroscopic gravity; and (6)~the effective temperature.

\begin{figure}
\begin{center}
\includegraphics[angle=-90, width=\hsize]{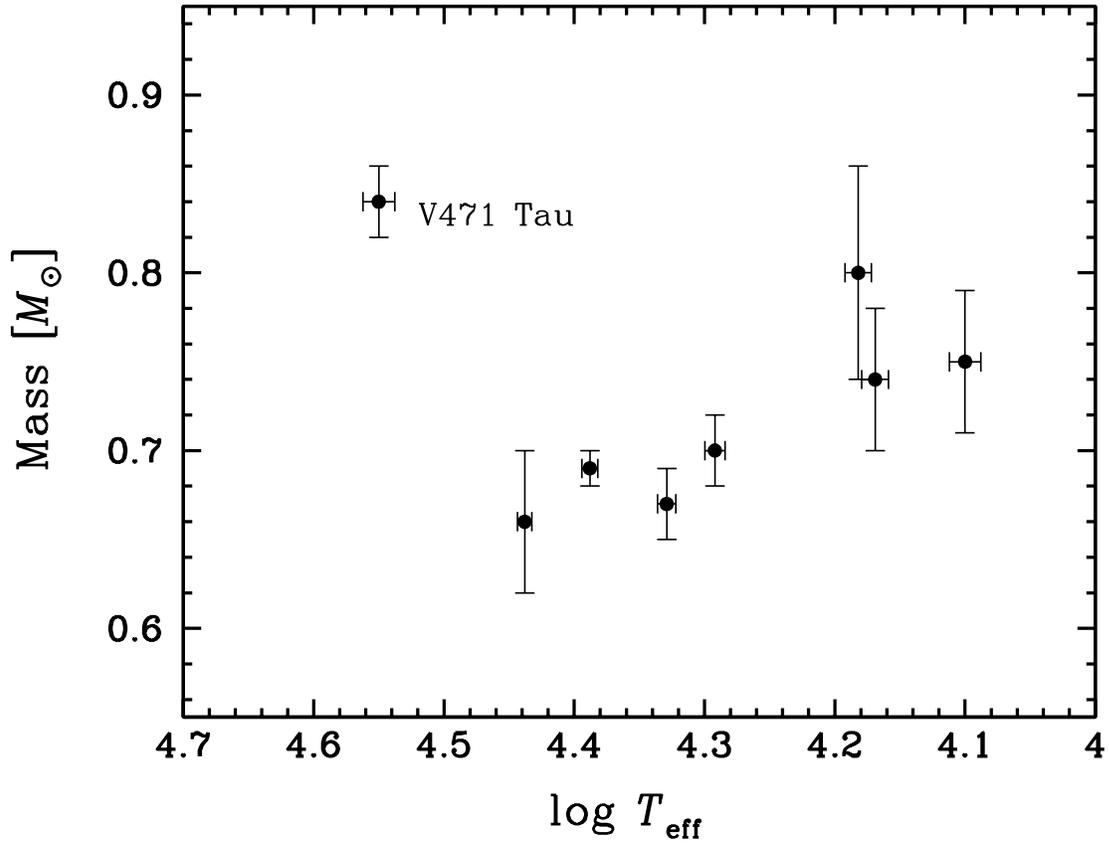}
\end{center}
\figcaption[f8.eps]{Positions of the single Hyades white dwarfs and of the
white dwarf in V471 Tau in the effective temperature vs.\ mass diagram. Data
are taken from Table~3;  gravitational-redshift masses are plotted for the
single white dwarfs, and our dynamical mass for V471 Tau. The single white
dwarfs show the expected increase in mass with decreasing temperature, but V471
Tau flagrantly violates the trend. 
\label{hyadeswd}}
\end{figure}

In Figure~\ref{hyadeswd} we plot the WD masses vs.\ effective temperatures.  We
have omitted HZ~9 from the figure, because, as just noted, the mass of its cool
companion must be assumed from its spectral type, and the pitfalls of this
assumption have been demonstrated vividly by V471~Tau.

Figure~\ref{hyadeswd} dramatically illustrates the paradox of the WD in
V471~Tau. We expect there to be a strong correlation between temperature and
mass, because the coolest WDs in the cluster should be the oldest ones; in
turn, these should be the most massive, since they are descended from the most
massive progenitor stars. And indeed such behavior is exhibited by all but one
of the Hyades WDs. V471~Tau flagrantly violates the correlation: it is both the
hottest WD in the Hyades {\it and\/} the most massive\footnote{Curiously, the
Praesepe cluster also contains a high-mass WD, LB~5893, that is anomalously hot
for its mass, as shown by Reid (1996), Claver, Liebert, \& Bergeron (1997), and
Claver et al.\ (2001); although LB~5893 is a single WD, it presents a paradox
similar to that of V471~Tau.}.  (Note that, if anything, the fact that V471~Tau
has emerged from a CE interaction makes the paradox even worse, since in the
absence of the binary interaction the WD's mass would have grown to an even
higher value.)


 
There are several possible astrophysical explanations for the high $\Teff$ 
value of the \VTau\ WD\null. We discuss the plausibility of each scenario in
the following subsubsections.

\subsubsection{Recent Star Formation}

It has been suggested (e.g., Eggen \& Iben 1988; see also Barstow et al.\ 1997,
their \S6.1) that the paradox of the hot WD in V471~Tau could be resolved if
the system really is much younger than other members of the Hyades, having been
formed in a late burst of star formation in the cluster. However, Perryman et
al.\ (1998), in their detailed study of the Hyades H-R diagram based on {\it
Hipparcos\/} parallaxes, found that the stars within 10~pc of the cluster
center can be fit (after removing known binary systems) with a single-age
isochrone with an age of $625\pm50$~Myr.  There is thus no independent support
for any recent star-forming activity in the cluster core.

\subsubsection{Accretion}

Can the WD heating arise from accretion from the stellar wind of the dK
component? The observed luminosity of the WD is $L_{\rm WD} =
0.15\,L_\sun$\null. The luminosity available from accretion at a rate of $\dot
M$ is
\begin{equation}
L_{\rm accr} \simeq 0.5 \, G M_{\rm WD} \dot M/R_{\rm WD} 
           \simeq 0.12 \Biggl(\frac{M_{\rm WD}}{0.8\,M_\sun}\Biggr)
                        \Biggl(\frac{R_{\rm WD}}{0.01\,R_\sun}\Biggr)^{-1}
		  \Biggl(\frac{\dot M}{10^{-10}\,M_\sun\rm\,yr^{-1}}\Biggr) \,
			  L_\sun \,  ,
\end{equation}
so that it would take an accretion rate of  
$\sim$$10^{-10}\,M_\sun\rm\,yr^{-1}$ to provide the luminosity. The entire
mass-loss rate from the dK star due to its stellar wind is probably smaller
than this (Mullan et al.\ 1989; Bond et al.\ 2001), and moreover we would
expect only a small fraction of the wind ($\approx$$10^{-2}$) to be captured by
the WD, even in the absence of magnetic effects.  In fact, however, our GHRS
observations of the photospheric \ion{Si}{3} line (Sion et al.\ 1998) indicate
that the actual accretion rate onto the WD is less than
$10^{-17}\,M_\sun\rm\,yr^{-1}$ (as supported also by the EUV observations of
Dupuis et al.\ 1997); we argued on this basis that most of the material that
attempts to accrete is rejected by a magnetic propeller effect.

Thus it appears that accretion luminosity is inadequate for explaining the high
temperature of the WD.

%

\subsubsection{Recent Nova}

Could the WD in V471~Tau still be hot following a recent nova outburst?

The possibility of a relatively recent nova event received some support when
Bruhweiler \& Sion (1986) and Sion et al.\ (1989) reported absorption lines
with large negative velocities ($-260$, $-590$, and $-1200\kms$) in \IUE\/
spectra of the binary; these authors suggested that the lines might arise in a
shell ejected during an ancient nova eruption, although for the two
lower-velocity components they could not rule out ejection events associated
with the K dwarf. They also pointed out that Chinese observers recorded a
possible nova in the year 396, lying near the location of V471~Tau in the sky
(cf.\ Pskovskii 1979 and references therein). Clark \& Stephenson (1977) place
the 396 event at the approximate position
$(\alpha,\delta)_{1950}=\rm4^h,+20^\circ$, as compared with $\rm3^h48^m,
+17^\circ06'$ for V471~Tau.  The brightness of the event was roughly consistent
with that of a DQ~Her-type classical-nova outburst at the distance of V471~Tau,
according to Hertzog (1986). On the other hand, Clark \& Stephenson state that
the event was ``at best a {\it probable\/} nova,'' and it remains possible that
the ancient report refers to a supernova, comet, or some other transient event
unrelated to V471~Tau, or to a nova outburst of a different object. Moreover,
the S/N of the claimed \IUE\/ detection of high-velocity lines was quite low,
and our GHRS observations reported here along with a preliminary examination of
more recent spectra obtained with the Space Telescope Imaging Spectrograph (to
be discussed in a separate paper) have not confirmed the existence of these
features. In addition, there is no evidence for an ejected nova shell in deep,
wide-angle H$\alpha$ direct imaging of the environment around V471~Tau
(J.~Gaustad, 1999 private communication).  

In a theoretical study, Prialnik \& Shara (1986) found that a $1.25\,M_\sun$ WD
will cool to the effective temperature of V471~Tau in $\la$$10^3$~yr following
a nova outburst, assuming that mass transfer from the cool companion drops to
zero following the explosion---a debatable proposition for classical novae, but
appropriate for V471~Tau. The probability of catching the system in an
immediate post-nova stage is thus extremely low. 

The post-nova cooling timescales are somewhat longer for lower-mass WDs, such
as that in V471 Tau. Theoretical studies indicate that a WD with the mass of
that in V471~Tau must accrete a hydrogen-rich envelope of
$\sim$$3\times10^{-4}\,M_\sun$ in order to initiate a thermonuclear runaway
(see Table~1 of Gehrz et al.\ 1998). The star's subsequent nuclear-burning
timescale (i.e., the time required to burn the envelope) would then be of order
$10^3$~yr (Gehrz et al.), and it would take somewhat longer to cool to the
observed temperature in V471~Tau. However, this still leaves the probability of
observing the system at the present epoch quite low.

If we were to suppose that the WD in V471~Tau was actually formed some
$5\times10^8$~yr ago, then it could have built an envelope mass of 
$3\times10^{-4}\,M_\sun$ with an accretion rate of $\dot M \simeq
6\times10^{-13} \, M_\sun \rm\,yr^{-1}$. Curiously, this value of $\dot M$ is
close to what the Bondi-Hoyle rate would be for a total mass-loss rate from the
K~dwarf of $\sim$$10^{-11}\, M_\sun \rm\,yr^{-1}$, which is the lower limit
reported by Mullan et al.\ (1989) from \IUE\/ absorption-line observations
obtained when the WD was just above the surface of the K star. However, our
GHRS observations of the WD (Sion et al.\ 1998) show that the magnetic
propeller mechanism actually limits the accretion rate to several orders of
magnitude lower than this; thus the WD would not be expected to have
accumulated enough hydrogen for even one nova outburst within the age of the
Hyades cluster. 

On the other hand, it is not impossible that the WD emerged from the CE event
with a hydrogen envelope already close to the mass necessary for a nova
outburst. Conceivably, also, the time-averaged accretion rate could be higher
than what it happened to be during our GHRS observations, due to episodic
flaring activity on the dK star.  In sum, it turns out to be surprisingly
difficult to rule out categorically that V471~Tau has had a nova outburst in
the recent past, although we deem this possibility to be unlikely from a
probabilistic standpoint.

\subsubsection{Blue-Straggler Scenarios}

The paradox of the hot WD in V471~Tau is, of course, reminiscent of the general
problem of blue stragglers in open and globular clusters. Blue stragglers are
stars on or near the MS, but lying above the MS turnoff and thus
apparently younger than the bulk of the cluster members. The possible
connection to V471~Tau is that a blue straggler should eventually evolve to
become an anomalously hot WD in the cluster. There is a general consensus that
many blue stragglers are merged binaries. However, a straightforward binary
merger scenario cannot explain V471~Tau, which contains a WD in a system that
is {\it still\/} a close but detached binary.

\citet{ibe99} have proposed a scenario that solves this problem by having
V471~Tau belong initially to  a {\it triple\/} system.  Unlike hierarchical
triples, whose orbits can remain stable for long intervals, the proposed
progenitor system in their scheme was among the $\sim$10--20\% of triples that
are non-hierarchical and was thus on the brink of instability.

In the Iben-Tutukov scenario, the initial system contained an Algol-type close
binary, while the \kd\ orbited the Algol at a larger distance. The WD
progenitor in the Algol system began with a mass similar to  (or perhaps
slightly less than) that of the present-day MS turnoff stars in the Hyades
($\sim$$2.5\,M_\sun$; Weidemann et al.\ 1992), and the hypothetical donor
companion had an initial mass somewhat above the MS turnoff. The donor evolved
to fills its Roche lobe, and conservative mass transfer ensued. As the mass
transfer proceeded it increased the mass of the WD progenitor. Once the mass
ratio was inverted, the orbital separation in the Algol binary would begin to
increase, until it exceeded the limit of stability. At that point, one of the
stars, most likely the one of lowest mass, which would now be the Algol donor,
was ejected from the system, leaving a binary composed of a MS primary, now a
blue straggler with a mass that could approach twice the turnoff mass, and the
K~companion.  This system would then evolve into a CE event as outlined above,
giving rise to a much closer pair consisting of the WD core of the Algol
accretor and the K~star.

Two possible objections to this scenario arise immediately, however. One is
that the system should have received a recoil velocity of order the orbital
speed (scaled appropriately by the masses) when one star was ejected, yet the
binary's space motion is measured to be within a few $\!\kms$ of the cluster
velocity. The other is that several authors (Beavers, Herczeg, \& Liu 1986;
\Ibanoglu\ et al.\ 1994; Guinan \& Ribas 2001) have proposed that V471~Tau has
a distant third companion, with an orbital period of $\sim$30~yr and
eccentricity $\sim$0.3. This suggestion is based on eclipse timings that show a
periodic variation in the orbital period; if attributed to the light
travel-time effect, the third companion's mass is only $M_3 \sin i_3 \simeq
0.04\,M_\sun$, according to Guinan \& Ribas. It would appear unlikely that such
a loosely bound third body would have survived the triple-star interaction
outlined above, without at least having acquired a very large orbital
eccentricity, although of course detailed modelling would be required to
support this statement.  Alternatively, one might argue that this third body
{\it is\/} the putative former donor object, although again one might expect a
larger eccentricity, and also in the Iben-Tutukov scenario the subgiant removed
from the Algol system has a mass about an order of magnitude larger than
inferred for the present-day distant companion. Actually, however, only about
one cycle of the proposed 30-yr orbit has been observed, so it may be premature
even to conclude that there is a third body (as opposed to intrinsic period
variability, which has been seen in many other close binaries).

In any case, we suggest that an alternative and simpler triple-star
scenario may account for all of the properties of V471~Tau without violating
any known constraints.   Suppose that the initial system consisted of a very
close pair of MS stars, with the original primary somewhat more massive than
the current MS turnoff and the original secondary somewhat less massive. The
current dK star would, again, be a more distant companion. A sufficiently close
MS pair will evolve into a ``Case~A'' Algol-type system when the primary
expands to fill its Roche lobe. As shown in the recent simulations by Nelson \&
Eggleton (2001), many Case~A systems evolve eventually into a contact
configuration. This will lead to a merger of the close pair and formation of a
blue straggler.  The straggler would subsequently evolve into an AGB star, at
which point it would fill the Roche lobe of its orbit with the more distant dK
star, enter into a CE interaction, eject the envelope as a planetary nebula,
and evolve into the present V471~Tau configuration.

Fortuitously, Nelson \& Eggleton (2001, their Fig.~4) present a simulation that
comes remarkably close to accounting for the evolution of the inner pair in
V471~Tau, under our Case~A scenario. They start with a ZAMS pair of masses 2.8
and $2.5\,M_\sun$ and an initial orbital period of 1.6~day. The original
primary in this binary commences Roche-lobe overflow at an age of 370~Myr, the
mass ratio quickly inverts, and the system becomes a contact binary at 490~Myr
when the original secondary evolves to fill its own Roche lobe. Just before
coming into contact, the masses are 2.0 and $3.3\,M_\sun$. The ensuing merger
would produce a blue straggler of $5.3\,M_\sun$, which would have a MS lifetime
of $\sim$100~Myr. Thus, at an age of $\sim$600~Myr (very close to  the age of
the Hyades), the merged object would evolve to the AGB stage, and enter into a
CE event with the putative more distant dK companion. The outcome would be a
very close pair, consisting of the dK MS star and a hot WD with a mass that
could approach that of a remnant of a single $5.3\,M_\sun$ star.  

Such a scenario is admittedly {\it ad hoc\/} and has few, if any,
observationally testable consequences, apart of course from the very existence
of the present-day V471~Tau system\footnote{The rotation speed of the V471~Tau
WD, $v \sin i \simeq 80\kms$, is near the high end of the range observed in
single WDs (see Koester et al.\ 1998) and might be thought to support a
binary-merger scenario, but a small amount of mass accretion from the bloated
dK component near the end of the CE interaction might also account for the WD's
high angular momentum}. We can, however, point out that catalogs of field
triple (or higher multiplicity) systems have been presented by Fekel (1981),
Chambliss (1992), and Tokovinin (1997). Among these multiples are some that
have parameters more or less similar to that of the proposed progenitor of
V471~Tau. For example, Algol ($\beta$~Persei) itself consists of an inner
eclipsing pair with a total mass of $\sim$$4.5\,M_\sun$, orbited by an outer F
dwarf with a period of 1.9~yr (see Fekel 1981, Nelson \& Eggleton 2001, and
references therein). The triple system of $\delta$~Librae is also quite similar
to Algol (Worek 2001), with an outer late G~dwarf orbiting the close eclipsing
pair in a 2.8-yr period. Hence it is clear that nature does occasionally create
real triples with properties similar to those proposed here.


%
%

\section{Implications for Common-Envelope Evolution}

As discussed in \S1, one of the motivations for this study was to determine
$\ace$, the efficiency with which orbital energy went into ejecting matter from
the V471~Tau system during its recent CE interaction, as defined in our
equation~(1).  Our initial expectation was that this determination would be
straightforward: because of the high temperature and short cooling age of
the WD, we could simply assume the total mass of the WD progenitor (hereafter
we refer to this object as ``the AGB star'') to have been equal to that of
Hyades MS turnoff stars. The AGB star's core mass would be taken to be the
current mass of the WD, and the radius and binding energy of the AGB star's
envelope could be calculated easily from stellar models. Then, knowing all
parameters of the post-CE binary, we could calculate $\ace$ entirely from known
quantities. (It should be noted that in this section we consider only the CE
interaction of the AGB star with its dK companion, apart from any previous
events that might have created or altered the AGB star, as discussed in the
previous section.) 

Unfortunately, however, the evolutionary paradox of the WD in V471~Tau forces
us to abandon the assumption that the AGB star had an initial mass equal to the
current Hyades turnoff mass.  Instead, it was very probably higher.

Although V471~Tau has now become a less ideal laboratory for CE evolution than
previously thought, we can still place useful limits on the value of $\ace$,
since the range of possible AGB star masses is fairly limited.   A recent
observational study of the WD initial-mass/final-mass relation (IMFMR) by
Jeffries (1997) indicates (for Hyades metallicity) that the WD mass of
$0.84\,M_\sun$ in V471~Tau corresponds to an initial MS mass of
$\sim$$4\,M_\sun$\null. From the older Weidemann (1987) IMFMR the initial mass
would have been higher, $\sim$$5.8\,M_\sun$. These values are, of course, lower
limits on the AGB star's mass in V471~Tau, since the CE evolution would if
anything have limited the growth of the core mass to a value below that
attainable by an undisturbed single star of the same initial mass. However, if
our Algol merger scenario is correct, the initial mass would have been limited
to about twice the current Hyades turnoff mass.

We have chosen to estimate $\ace$ for a range of assumed AGB star masses from
3.5 to $5.5\,M_\sun$, with the low end extended slightly below the Jeffries
lower limit to allow for the possibility of some stellar-wind mass loss prior
to the CE event, and the upper limit corresponding to slightly more than twice
the turnoff mass. We determine the luminosity of the AGB star at the moment of
Roche-lobe overflow by taking $M_{\rm core}$ to be the current WD mass of
$0.84\,M_\sun$, and using the theoretical AGB core-mass/luminosity relation of
Vassiliadis \& Wood (1994):
\begin{equation}
L_{\rm AGB}/L_\sun = 
  56694 \, [(M_{\rm core}/M_\sun) - 0.5] \, ,
\end{equation}
which implies an AGB luminosity of $ L_{\rm AGB} = 19275\,L_\sun$\null.  For
the radius of the AGB star at the beginning of the CE interaction we use a
formula based on Hurley, Pols, \& Tout (2000):
\begin{equation}
R_{\rm AGB}/R_\sun = 1.218 \, M_{\rm AGB}^{-0.34} \, 
  [({L_{\rm AGB}/L_\sun})^{0.41} + 0.335 ({L_{\rm AGB}/L_\sun})^{0.75}] \, ,
\end{equation}
which applies for Hyades metallicity, as set forth in detail by Hurley et al.
We can thus calculate the radius of the AGB star and equate it to the radius
of its Roche lobe, $R_L$, at the onset of mass transfer to the dK companion.
Next, for each assumed value of $M_{\rm AGB}$, we can calculate the initial
orbital separation, $a_i$, from the Eggleton (1983) formula
\begin{equation}
R_L = \frac{0.49\,q^{2/3}\,a_i}{0.6\,q^{2/3} + \ln(1+q^{1/3})} \, ,
\end{equation}
where $q=M_{\rm AGB}/M_2$ and $M_2$ is the mass of the companion. To proceed,
we need to calculate the binding energy of the AGB star's envelope at the
beginning of the CE event and the change in orbital energy (potential plus
kinetic) from the beginning to the end of the event.

The binding energy can be calculated explicitly from stellar models. However,
as discussed by, e.g., Han, Podsiadlowski, \& Eggleton (1995) and Livio (1996),
in practice two significantly different approximations have been used in the
literature:

(1).~de~Kool (1990) writes the binding energy, $\Delta E_{\rm bind}$, of the
AGB star's envelope as
\begin{equation}
\Delta E_{\rm bind} = \frac{G M_{\rm AGB} M_{\rm env}}{\lambda R_{\rm AGB}} \, ,
\end{equation}
where $M_{\rm env}$ is the envelope mass of the AGB star, given by $M_{\rm env}
= M_{\rm AGB} - M_{\rm core}$, and $\lambda$ is a constant of order 0.5,
according to de~Kool's comparison with detailed stellar models.  This
formulation neglects the (relatively small) contribution to the gravitational
potential well from the companion star. It also assumes that the envelope
material is ejected at just the escape velocity.

The CE interaction is assumed to eject $M_{\rm env}$ from the system, leaving a
much closer binary composed of the AGB star's core and the secondary star. In
de~Kool's formulation, the change in the orbital energy as a result of the
interaction is given by
\begin{equation}
\Delta E_{\rm orb} = \frac{G M_{\rm core} M_2}{2a_f} -
  \frac{G M_{\rm AGB} M_2}{2 a_i} \, ,
\end{equation}
where $a_f$ is the final orbital separation and $M_2$ is assumed to be
unchanged during the interaction.

Adopting the remaining observational quantities of $M_2=0.93\,M_\sun$ and 
$a_f=3.3\,R_\sun$, we now have all of the information needed to calculate
$\ace$ as a function of assumed $M_{\rm AGB}$, using equations~(1) and
(16)--(20). The results are presented in Table~\ref{acetable}, in which the
calculated efficiency parameter is denoted $\alpha_{\rm CE,\,de Kool}$. 

Table~\ref{acetable} indicates that the V471~Tau progenitor system had an
orbital period of several years, with the AGB star having a radius of about
half the orbital separation. The calculated de~Kool $\ace$ lies between about
0.3 and 1, depending on the assumed mass of the AGB star. Recent detailed
hydrodynamical calculations of CE interactions for systems with parameters 
similar to those that are likely for the V471~Tau progenitor (Yorke,
Bodenheimer, \& Taam 1995; Sandquist et al.\ 1998) have yielded values of
$\ace$ of $\sim$0.3--0.6, in good agreement with our empirical results.


(2).~In a separate series of papers, Tutukov, Yungelson, Iben, Livio, and
collaborators have used a different pair of equations to approximate the
binding and orbital energies (see the review by Iben \& Livio 1993). They write
the envelope's binding energy as
\begin{equation}
\Delta E_{\rm bind} = \frac{G (M_{\rm AGB}+M_2) M_{\rm env}}{2 a_i} \, ,
\end{equation}
i.e., they consider the stage at which the CE has already grown to a radius of
$a_i$, and they include the contribution of the secondary star to the potential
as well.  A smaller difference is that these
collaborators write the change in orbital energy as
\begin{equation}
\Delta E_{\rm orb} = \frac{G M_{\rm core} M_2}{2a_f} -
  \frac{G M_{\rm core} M_2}{2 a_i} \, ,
\end{equation}
that is, they consider only the change in the orbital energy of the degenerate
core and the MS companion, and not the additional (but small) contribution from
the orbital energy of the AGB envelope.  We calculated the CE efficiency for
V471~Tau using equations~(21)--(22) instead of (19)--(20), and tabulate the
results, denoted $\alpha_{\rm CE,\,Tutukov}$, in the last column of Table~4. In
this case, the efficiency parameter is found to be  of order 0.1.  As pointed
out by Han et al.\ (1995) and Livio (1996), $\Delta E_{\rm bind}$ calculated in
this fashion is significantly smaller than calculated using de~Kool's
approximation---by a factor of $\sim$6 in the case of V471~Tau; thus the
derived $\ace$ is also smaller by the same factor. 

\section{Summary and Discussion}

This paper reports \HST/GHRS spectroscopy at \Lya\ of the  hot
component in the eclipsing DA+dK pre-cataclysmic binary \VTau. The
\Lya\ radial velocities of the WD, combined with ground-based
measurements of eclipse timings and of the radial velocities and rotational
velocity of the dK star, constrain the orbital inclination to be $\sin i =
0.976$. The resulting dynamical masses for the components are $M_{\rm WD}=0.84$
and $M_{\rm dK}=0.93\,M_\sun$. Both values are significantly larger than
adopted in previous investigations, which had to rely (erroneously) on an
assumed mass for the K star based on its spectral type.

The dK component is remarkable in being about 18\% larger than a normal Hyades
dwarf of the same mass. We attribute the expansion to the large ($\sim$25\%)
fractional coverage of the star's surface by starspots. In response to this
covering, the star has expanded in order to radiate the luminosity generated in
its core.

Although such an expansion is predicted theoretically, our observations are
probably the clearest demonstration yet that the phenomenon does exist. The
oversized radius will have important implications for the future evolution of
V471~Tau and other pre-cataclysmic binaries, if it is a general property of
such systems. For example, V471~Tau will become a cataclysmic variable earlier
than otherwise expected, since it will fill its Roche lobe earlier.  Moreover
it is probably incorrect to assume a normal main-sequence relationship between
mass, radius, and spectral type for the cool components of cataclysmic and
pre-cataclysmic binaries (cf. Smith \& Dhillon 1998).

The effective temperature of the WD was determined by model-atmosphere fitting
to the \Lya\ profile, and was found to be 34,500~K\null. The radius
of the WD, $0.0107\,R_\sun$, was determined from a radiometric  analysis  and
from eclipse ingress timings.  Availability of direct measurements of the
radius and mass of the star allow the most stringent comparisons yet with the
theory of WD structure. The position of the star in the mass-radius plane is
shown to be in full accord with the predictions of theory for a degenerate
carbon-oxygen WD with a surface temperature equal to that observed. The
position of the WD in the H-R diagram is also fully consistent with that
expected for a WD with our dynamically measured mass.  The inferred cooling age
of the WD is close to $10^7$~yr.

The high effective temperature, high mass, and apparent short cooling age of
the WD present an evolutionary paradox. The WD is the most massive known in
the Hyades, but also the hottest, in direct conflict with theoretical
expectation. We examined possible resolutions of the paradox, and concluded that
the WD is probably indeed very young (rather than being anomalously hot through
some more exotic mechanism, such as a recent nova outburst). Most probably it
is descended from a blue straggler. A plausible scenario is that the progenitor
system was a triple, with a close inner pair of main-sequence stars whose
masses were both similar to that of the present cluster turnoff. These stars
first became an Algol-type binary, which merged after several hundred million
years to produce a single star of about twice the turnoff mass. When this star
evolved to the AGB phase, it underwent a common-envelope interaction with a
distant dK companion, spiralled down to its present separation, and ejected the
envelope. We estimate that the common-envelope efficiency parameter, $\ace$,
was of order 0.3--1.0, in good agreement with hydrodynamical simulations. 

In the widely quoted approximation used by Tutukov and collaborators, however,
our estimated $\ace$ is of order 0.1.  A value this low makes it easier to
produce short-period post-CE systems, including CVs and double-degenerate
binaries that may be Type~Ia supernova progenitors.  A specific, and testable,
prediction from the population syntheses of Yungelson et al. (1993) is that,
with $\ace\simeq0.1$, the orbital periods of close-binary nuclei of planetary
nebulae will typically range from 0.3 to 30~days.

The Space Telescope Imaging Spectrograph (STIS) is now available onboard \HST,
and covers the entire UV spectrum rather than a small window around
\Lya. We are currently undertaking followup observations of \VTau\
using STIS, and have also obtained STIS spectra of HZ~9, the other known
pre-cataclysmic binary in the Hyades. We hope that these data  will provide
even more accurate radial velocities (through observations of photospheric
metallic lines), including a direct measurement of the gravitational redshift.
These observations should allow us to test, refine, and expand upon the results
reported here for this endlessly fascinating binary system.



\acknowledgments

This paper had a long gestation period, during which we benefitted from the
assistance of many people. The initial concepts for the observing program arose
from discussions between HEB and John Stauffer. The demanding \HST\/ scheduling
was accomplished by many talented people at STScI, including Dustin Manning and
Denise Taylor. Lengthy initial assessments of the data and preliminary analyses
were carried out by Karen Schaefer and Rex Saffer. We had useful discussions
with, among others, Fuhua Cheng, Chris Clemens, Jean Dupuis, Ed Guinan, Colleen
Henry, Steve Hulbert, Icko Iben, Steve Kawaler, Jim Liebert, Mario Livio,
Dermott Mullan, Dina Prialnik, Neill Reid, and Matt Wood. Ivan Hubeny very
kindly provided his atmosphere code and much assistance and advice in running
it. John Gaustad generously obtained wide-field H$\alpha$ images of the
V471~Tau field and provided them in advance of publication. The referee,
Stephane Vennes, made several comments that helped us improve the paper.

We acknowledge financial support from STScI grant GO-5468.  The work of EMS and
CH was also supported by NSF grant AST99-09155 to Villanova University and by
NASA ADP grant  NAG5-8388. 


\clearpage

\clearpage

\clearpage

\clearpage

\clearpage

\clearpage

\clearpage

\clearpage


\clearpage

\begin{deluxetable}{llcccc}
\tablewidth{0pt}
\tablecaption{V471~Tau GHRS Spectra and Radial Velocities\label{spec_tab}}
\tablecolumns{2}
\tablehead{
\colhead{Dataset} & \colhead{Observation} & \colhead{HJD} &
\colhead{Orbital Phase}  & \colhead{$V_{\rm WD}$}  & 
\colhead{$\sigma_{V}$} \\
\colhead{Name}   & \colhead{Date (UT)}     & \colhead{(mid-exposure)} &
\colhead{(mid-exposure)} & \colhead{($\!\kms$)} & 
\colhead{($\!\kms$)}}
\startdata
Z1KJ0403T$^{\rm a}$ & 1993 October 6  & 2449267.10786 & 0.363 & $-139$ & 17\\
Z1KJ0203P$^{\rm a}$ & 1993 October 8  & 2449269.18274 & 0.344 & $-161$ & 18\\
Z1KJ0803T$^{\rm a}$ & 1993 October 8  & 2449269.38805 & 0.738 & $+167$ & 17\\
Z1KJ0603T$^{\rm a}$ & 1993 October 10 & 2449270.98981 & 0.811 & $+127$ & 17\\
Z2I70G03T 	    & 1994 October 17 & 2449643.09303 & 0.769 & $+224$ & 10\\
Z2I70803T 	    & 1994 October 18 & 2449644.36700 & 0.214 & $-108$ & 9\\
Z2I70E03T 	    & 1994 October 19 & 2449645.17175 & 0.758 & $+210$ & 11\\
Z2I70A03T 	    & 1994 October 20 & 2449646.24396 & 0.815 & $+181$ & 10\\
Z2I70403T 	    & 1994 October 21 & 2449646.98247 & 0.232 & $-111$ & 8\\
Z2I70C03T 	    & 1994 October 23 & 2449649.32981 & 0.736 & $+207$ & 10\\
Z2I70603T 	    & 1994 October 25 & 2449651.14045 & 0.210 & $-122$ & 9\\
Z2I75203T 	    & 1995 March 16   & 2449792.89900 & 0.203 & $-115$ & 10\\
\enddata
\tablenotetext{\rm a}{Pre-COSTAR observations, STScI Program ID 4344; 
all others are post-COSTAR, STScI Program ID 5468.}
\end{deluxetable}

\begin{deluxetable}{lll}
\tablewidth{0pt}
\tablecaption{Measured and Derived V471~Tauri System Parameters\label{param_tab}}
\tablehead{
\colhead{Parameter} & \colhead{Value} & \colhead{Source$^{\rm a}$ } }
\startdata
Parallax           & $21.00\pm0.40$ mas & de Bruijne et al.\ 2001 \\
Distance           & $47.6 \pm 0.9$ pc  & de Bruijne et al.\ 2001 \\
$P_{\rm orb}$      & 0.52118373 day & \S2 \\
$T_{\rm surf,dK}$  & $5,040 \pm 100$ K & \S5.1 \\
$T_{\rm eff,WD}$   & $34,500 \pm 1,000$ K & \S3.2 \\
$\log g_{\rm WD}$  & $8.31 \pm 0.06$ & \S7.4 \\
$v_{\rm rot,dK} \sin i$ & $91 \pm 4\kms$ & Ramseyer et al.\ 1995 \\
$K_{\rm dK}$  & $148.46 \pm 0.56 \kms$ & Bois et al. 1988 \\
$K_{\rm WD}$  & $163.6 \pm 3.5 \kms$  & \S4 \\
$\sin i$ & $0.976 \pm 0.020$ & \S5.1 \\
$a$ & $3.30 \pm 0.08 \, R_\sun$ & \S5.2 \\
$R_{\rm dK}$  & $0.96 \pm 0.04 \, R_\sun$ &  \S5.1 \\
$R_{\rm WD}$  & $0.0107 \pm 0.0007 \, R_\sun$ & \S7.2 \\
$M_{\rm dK}$ & $0.93 \pm 0.07 \, M_\sun$ & \S5.2 \\
$M_{\rm WD}$ & $0.84 \pm 0.05 \, M_\sun$ & \S5.2 \\
$L_{\rm dK}$ & $0.40 \pm 0.02 \, L_\sun$ & \S6 \\
$L_{\rm WD}$ & $0.146 \pm 0.016 \, L_\sun$ & \S7.3 \\
$t_{\rm cool, WD}$ & $10^7$ yr & \S7.3 \\
\enddata
\tablenotetext{\rm a}{Sections in this work, unless otherwise indicated.}
\end{deluxetable}

%


\begin{deluxetable}{lccccc}
\tablewidth{0pt}
\tablecaption{Properties of the Hyades White Dwarfs\label{hyad_tab}}
\tablehead{
\colhead{Name} & \colhead{WD} & \colhead{$M_{\rm dyn}/M_{\odot}^{\rm a}$} 
& \colhead{$M_{\rm GR}/M_{\odot}^{\rm b}$}
& \colhead{$M_{\rm sp}/M_{\odot}^{\rm c}$}
& \colhead{$\Teff$ (K)$^{\rm d}$}
}
\startdata
\VTau\   & 0347+171 & $0.84\pm0.05$& $\dots$ &	$\dots$   & $34,500\pm1,000$  \\
HZ 4     & 0352+096 & $\dots$ & $0.74\pm0.04$ & $0.72\pm0.03$ & $14,770\pm350$ \\
LB 227   & 0406+169 & $\dots$ & $0.80\pm0.06$ & $0.80\pm0.04$ & $15,190\pm350$ \\
VR 7     & 0421+162 & $\dots$ & $0.70\pm0.02$ & $0.68\pm0.03$ & $19,570\pm350$ \\
VR 16    & 0425+168 & $\dots$ & $0.69\pm0.01$ & $0.68\pm0.03$ & $24,420\pm350$ \\
HZ 9     & 0429+176 &  $0.51\pm0.10$& $\dots$ & $\dots$      & $20,000\pm2,000$ \\
HZ 7     & 0431+125 & $\dots$ & $0.67\pm0.02$ & $0.65\pm0.03$ & $21,340\pm350$ \\
LP 475-242 & 0437+138 & $\dots$ & $0.75\pm0.04$ & $\dots$    & $12,600$    \\
HZ 14    & 0438+108 & $\dots$ & $0.66\pm0.04$ & $0.68\pm0.03$ & $27,390\pm350$ \\
%
\enddata
\tablenotetext{\rm a}{Dynamical masses. Sources: V471~Tau: this paper; HZ~9:
\citet{sta87}.}
\tablenotetext{\rm b}{Gravitational-redshift (GR) masses from \citet{rei96}.}
\tablenotetext{\rm c}{Spectroscopic masses from \citet{rei96} if available, or
from \citet{ber95}.}
\tablenotetext{\rm d}{Effective temperatures. Sources: V471~Tau: this paper;
HZ~9: Guinan \& Sion (1984); LP~475-242: \citet{rei96}, no error quoted;  other
stars: \citet{ber95}.}
\end{deluxetable}

\begin{deluxetable}{ccccccc}
\tablewidth{0pt}
\tablecaption{Calculations of Common-Envelope Efficiency, $\ace$\label{acetable}}
\label{acetable}
\tablehead{
\colhead{$M_{\rm AGB}$} & \colhead{$q$} & \colhead{$R_{\rm AGB}$} &
\colhead{$a_i$} & \colhead{$P_{{\rm orb},\,i}$} &
\colhead{$\alpha_{\rm CE,\,de Kool}$} & \colhead{$\alpha_{\rm CE,\,Tutukov}$}\\
\colhead{($M_\sun$)} & \colhead{} & \colhead{($R_\sun$)} & \colhead{($R_\sun$)}&
\colhead{(yr)} & \colhead{}& \colhead{}
}
\startdata
3.5	  &	 3.76    & 486 & 979  & 4.6  & 0.33 & 0.05 \\
4.0	  &	 4.30    & 464 & 915  & 3.9  & 0.47 & 0.07 \\
4.5	  &	 4.84    & 446 & 862  & 3.4  & 0.64 & 0.10 \\
5.0	  &	 5.38    & 431 & 818  & 3.0  & 0.84 & 0.13 \\
5.5	  &	 5.91    & 417 & 780  & 2.7  & 1.07 & 0.16 \\
\enddata
\tablenotetext{~}{Notes: Col.~1: assumed total mass of the AGB star in the
V471~Tau system at the onset of  the common-envelope (CE) interaction; its core
mass is taken to be $0.84\,M_\sun$. Col.~2: mass ratio relative to the
$0.93\,M_\sun$ dK secondary star. Col.~3: radius of the AGB star at CE onset.
Cols.~4 \& 5:  orbital separation and period of the AGB and dK stars at the
onset of the CE event. Cols.~6 \& 7:  the efficiency of conversion of orbital
energy into ejecting matter from the system, calculated as described in the
text according to the formulations of de~Kool and of Tutukov and collaborators,
respectively.}
\end{deluxetable}

\end{document}